\documentstyle[12pt,aaspp4,mystyle,flushrt]{article} 

\newcommand{\nc}{\newcommand}
\nc{\bec}{\begin{center}}
\nc{\enc}{\end{center}}
\nc{\beq}{\begin{equation}}
\nc{\enq}{\end{equation}}
\nc{\beqar}{\begin{eqnarray}}
\nc{\enqar}{\end{eqnarray}}
\nc{\bei}{\begin{itemize}}
\nc{\eni}{\end{itemize}}
\nc{\bee}{\begin{enumerate}}
\nc{\ene}{\end{enumerate}}

\nc{\lig}{light curve}
\nc{\namely}{{\it viz.}}
\nc{\ea}{et al.\ }

\nc{\pl}{period--luminosity}
\nc{\plr}{period--luminosity relation}
\nc{\plc}{period--luminosity--color}
\nc{\pca}{period--color--amplitude}

\nc{\rv}{R_V}
\nc{\av}{A_V}
\nc{\ai}{A_I}
\nc{\lp}{\log (P)}
\nc{\teff}{T_{\mathrm{eff}}}
\nc{\gteff}{\mbox{$ \log \teff \,$--$\,\log g $}}
\nc{\chisq}{\chi ^2}
\nc{\msun}{${\mathcal M}_\odot$}
\nc{\ksm}{km\,s$^{-1}$\,Mpc$^{-1}$}

\nc{\BV}{(B-V)}
\nc{\UB}{(U-B)}
\nc{\vi}{(V-I)}
\nc{\eBV}{E(B-V)}
\nc{\eUB}{E(U-B)}
\nc{\evi}{E(V-I)}
\nc{\cBV}{(B-V)_0}
\nc{\cUB}{(U-B)_0}
\nc{\cvi}{(V-I)_0}

\nc{\vmax}{V_{\mathrm{max}}}
\nc{\vmin}{V_{\mathrm{min}}}
\nc{\uncv}{\langle V\rangle }
\nc{\unci}{\langle  I\rangle }
\nc{\cv}{\langle  V\rangle _0}
\nc{\ci}{\langle  I\rangle _0}
\nc{\uncBVav}{\langle  B-V\rangle }
\nc{\uncviav}{\langle  V-I\rangle }
\nc{\viav}{\langle  V-I\rangle _0}
\nc{\BVav}{\langle  B-V\rangle _0}
\nc{\UBav}{\langle  U-B\rangle _0}
\nc{\uncvimax}{(V-I)|_{\mathrm{at}\: V_{\mathrm{max}}}}
\nc{\uncvimin}{(V-I)|_{\mathrm{at}\: V_{\mathrm{min}}}}
\nc{\vimax}{(V-I)_0|_{\mathrm{at}\: V_{\mathrm{max}}}}
\nc{\vimin}{(V-I)_0|_{\mathrm{at}\: V_{\mathrm{min}}}}
\nc{\viamp}{\Delta (V-I)}
\nc{\vamp}{\Delta V}
\nc{\visyn}{\vi_{\mathrm{synth}}}

\nc{\phiz}{\phi_0}
\nc{\phimin}{\phi_{\mathrm{min}}}
\nc{\phimax}{\phi_{\mathrm{max}}}
\nc{\phinew}{\phi_{\mathrm{n}}}
\nc{\phitilnew}{\phi'}

\lefthead{Mazumdar \& Narasimha}
\righthead{Galactic Cepheids and the Distance Scale}

\begin{document} 

\title{Some Characteristics of Galactic Cepheids\\
       Relevant to the Calibration of the Distance Scale}

\author{Anwesh Mazumdar\altaffilmark{1}
         and D. Narasimha\altaffilmark{2}}
\affil{Department of Astronomy and Astrophysics,
       Tata Institute of Fundamental Research, Homi Bhabha Road,
       Mumbai 400005, India.}
\altaffiltext{1}{e-mail: anwesh@astro.tifr.res.in}
\altaffiltext{2}{e-mail: dna@astro.tifr.res.in}

\begin{abstract}
An analysis of the observed characteristics of the Galactic Cepheid variables
is carried out in the framework of their \plr\
being used as a standard candle for the  distance measurement.
The variation of the observed number density of Galactic Cepheids as function 
of their period and amplitude along with stellar pulsation characteristics
is used to divide the population into two groups: one with low periods,
probably multi-mode or higher mode oscillators, and another of high period
variables which should be dominantly fundamental mode radial pulsators.
Methods to obtain extinction-corrected colors from multi-wavelength 
observations of the second group of variables are described and templates 
of the $\vi $ \lig s are obtained from the $V$ \lig s.
Colors computed from the model atmospheres are compared with the 
extinction-corrected colors to determine the Cepheid instability strip in 
the {\em mean surface gravity--effective temperature diagram}, and relations 
are derived between mean colors {\em $\BV$ vs period of pulsation}, {\em $\vi$ 
vs period}, and {\em $\vi$ at the brightest phase vs amplitude of pulsation}. 
The strength of the $\kappa$-mechanism in the envelope models is used
to estimate the metal dependency of the instability strip from which
an idea of the sensitivity of the \plr\ to the helium and
metal abundance is given. Some estimate of the mass of
Cepheids along the instability strip is provided.
\end{abstract}


\section{Introduction}
\label{sec:intro}

The classical Cepheid variables provide an important standard candle to 
measure distances to galaxies up to $\sim 30$ Mpc. The Cepheid Distance 
Scale is considered to be among
the most reliable methods because the physics of Cepheid pulsation is 
well-understood and the relation between the pulsation period and
luminosity of the star is observationally well-established.
The Cepheids are luminous, have a narrow range of surface temperatures;
their pulsation is very stable and exhibit large amplitude. 
The intrinsic scatter in their \plr\ is believed to
be only around $0.3$ mag.
However, the Cepheid distance scale
cannot be directly calibrated from the observation of nearby stars
and consequently, several systematic effects
still undermine its effectiveness as a standard primary candle
to determine extragalactic distances beyond a few Mpc. 
Some of the questions which have direct bearing on the problem of distance
calibration, but whose answers remain inconclusive in spite of extensive
research, are listed below.
\bei
\item   
Are the preferential pulsation modes of Cepheids period-dependent?
\item   
Is a single \plr\ applicable to the entire instability strip?
\item  
Is the \plr\ modified appreciably due to metallicity dependency of the 
stellar structure?
\item  
Is the \plc\ relation a better indicator of distance than the \plr\ only?
\eni

\noindent
Iben and Tuggle (1975) numerically computed the period and
luminosity of Cepheids for a range of masses and obtained a relation between
metallicity, surface temperature, period and luminosity. 
The \plc\ relation is found to be dependent on metallicity due to extreme
sensitivity of the color--temperature relation on chemical composition.
However, according to Becker, Iben and Tuggle (1977), within the
uncertainties, the relation between period and luminosity for the first and 
second crossings of the Cepheid instability strip by a particular star
does not crucially depend on the chemical composition. Since the time spent
in traversing the strip is largest for the second crossing, most of the 
observed Cepheids are in this stage of evolution. So, although conversion of
the period into a $V$-magnitude will introduce small effects due to surface
temperature and metallicity, a \plr\ derived from observations should
not be affected by small variations in the chemical composition. 
Indeed, the robustness of the \plr\ against changes in the chemical 
composition is borne out by theoretical as well as observational studies.
The theoretical models of Bressan \ea (1993) produce the same period (within
an error of $2 \% $) for a given luminosity, irrespective of the chemical
composition. From observations of Cepheids in M31, it appears that there is
no significant dependence of the \pl\ zero point on metallicity gradients
(\cite{fm:90}).
The recent review on the metallicity dependence of the Cepheid Distance
Scale in the context of the HST Key Project on Extragalactic Distance Scale
(\cite{kenn:98}) also leads to the same conclusion.
Similarly, even though color-color diagram
of Cepheids can be used to determine their metallicity, it is not an
improvement in terms of its application to the estimation of distances,
given that after extinction correction the color has larger error than
the $V$-magnitude (e.g., \cite{fg:93}). Clearly, in order to address the
above question, systematic work is required on 
the pulsation properties, evolution as well as  stellar atmospheric
structure, taking into account the onset of convection in the
atmosphere during the pulsation cycle.

In the present study we shall adopt the working hypothesis that the 
luminosity of the star {\it as a function of period} is not directly altered
by metallicity (subject to the star being a classical Cepheid),
and that the color of the star provides a better diagnostic for
the estimation of extinction than for the determination of the distance.
We devise methods to determine the extinction by the interstellar medium, 
and particularly emphasize the importance of observations in multi-wavelength 
bands, and also address the question of pulsation modes of Cepheids.

The Cepheids in our own Milky Way
galaxy have been observed by several astronomers over the years, and it is
possible to obtain multi-wavelength data as well as accurate periods for a
large number of them. A careful analysis of these Galactic Cepheids will
naturally provide a useful template for identifying and estimating the various
errors in the calibration of the Cepheid Distance Scale. A robust calibration
of this distance scale is particularly important for extending it to
extragalactic domains, as Cepheids are being observed in several far away
galaxies, including those in the Virgo Cluster, by the Hubble Space Telescope.
The hope is that 
distances based on these observations will ultimately lead to an accurate
determination of the Hubble Constant. In this context, we attempt to
provide a new calibration of the Cepheid Distance Scale, which is free from
many of the systematic errors. In an accompanying communication (which
we will refer to as Paper II), we
apply these results for estimating the distance to the Virgo Cluster, based on
the HST data for the Cepheids in the spiral M100.

This paper is organized as follows. 
In Section~\ref{sec:number} we discuss the number distribution of Cepheids
against their periods. In Section~\ref{sec:ligcur}, we demonstrate the
feasibility of obtaining accurate $\vi $ \lig s from the $V$
\lig\ of a Cepheid and limited number of observations in the I band.
This method is particularly useful for analyzing Cepheid data with few
observations in one band (as in HST observations). 
In Section~\ref{sec:extcor}, we
devise a formalism for extinction correction for each individual Cepheid,
based on model atmospheres and an $\rv$-dependent extinction law. Several
useful period--color and amplitude--color relationships are also derived.
Section~\ref{sec:modes} is concerned about the different modes of pulsation of 
Cepheid variables and their manifestations in the observed properties 
like period and amplitude of pulsation. We argue that it is necessary to choose
the correct lower cutoff period in the \plr\ in order
to prevent contamination from multi-mode pulsators. 
In Section~\ref{sec:mass}, we give
an estimate for Cepheid masses at different periods, based on our results
about the instability strip in the surface gravity versus effective
temperature plane. 
Some discussion on the metallicity effects are presented in 
Section~\ref{sec:metal} and the major limitations of the present work are 
listed in Section~\ref{sec:limit}.
The main conclusions from this work are summarized in Section~\ref{sec:concl}.

\section{Number Distribution of Galactic Cepheids}
\label{sec:number}

The classical Cepheid variables are radially pulsating giants and
supergiants, having pulsation periods in the range of less than a day to
upwards of 100 days. Their amplitude of light variation in the V (Johnson) 
band may be up to nearly 2 magnitudes although most of the Cepheids have
amplitude between 0.6 to 1.3 magnitude.
The General Catalogue of Variable Stars (GCVS) (\cite{khol:88}) provides a 
nearly exhaustive list of all the Cepheids observed in the Milky Way Galaxy 
along with their periods and $V$ amplitudes. 
Our principal source of photometric data for the Galactic Cepheids was in the
electronic form from the McMaster Cepheid Photometry and Radial Velocity
Data Archive, which contains detailed Julian Day versus magnitude
data in multiple wavelengths from different observers. Among the various
sources available at this website, we have chosen to
use data mainly from Coulson \& Caldwell (1985), Coulson, Caldwell \&
Gieren (1985), and in some cases from Berdnikov (1992), and Berdnikov
\& Turner (1995). The former two sources were preferred because they
provided homogeneous data in UBV (Johnson) and I (Cousins) passbands.
The latter sources were used to supplement data in U, B and I bands. 
However, our choice of stars is severely limited because we require 
data in all these four filters without internal inconsistencies in order
to carry out extinction correction.

From the GCVS, a  
distribution of nearly all classical Cepheids in the Galaxy, over a period 
range of 2 to 65 days is obtained. We have displayed the moving
averaged number density of Cepheids against the logarithm of their periods
in days ($\lp$) in Figure~\ref{fig:nod}. It
is evident that the majority of Cepheids observed in our galaxy have periods
less than 10 days. There is a small dip in the number density around a period
of 8--15 days. We fit a Gaussian curve to the number distribution
pattern at low periods, and find that the distribution function may be
split into two components. Below a period of 8 days, many of the Cepheids are
multi-mode pulsators, 
in which the fundamental mode may not be the dominant one. 
These Cepheids occupy the main peak in Figure~\ref{fig:nod}. Above
period of around 15 days, nearly all the Cepheids are single mode
pulsators. (The issue of modes of pulsation is discussed in detail in 
Section~\ref{sec:modes}).
In the intermediate range of periods, there exists a transition 
zone where not many Cepheids are observed. 

\figone{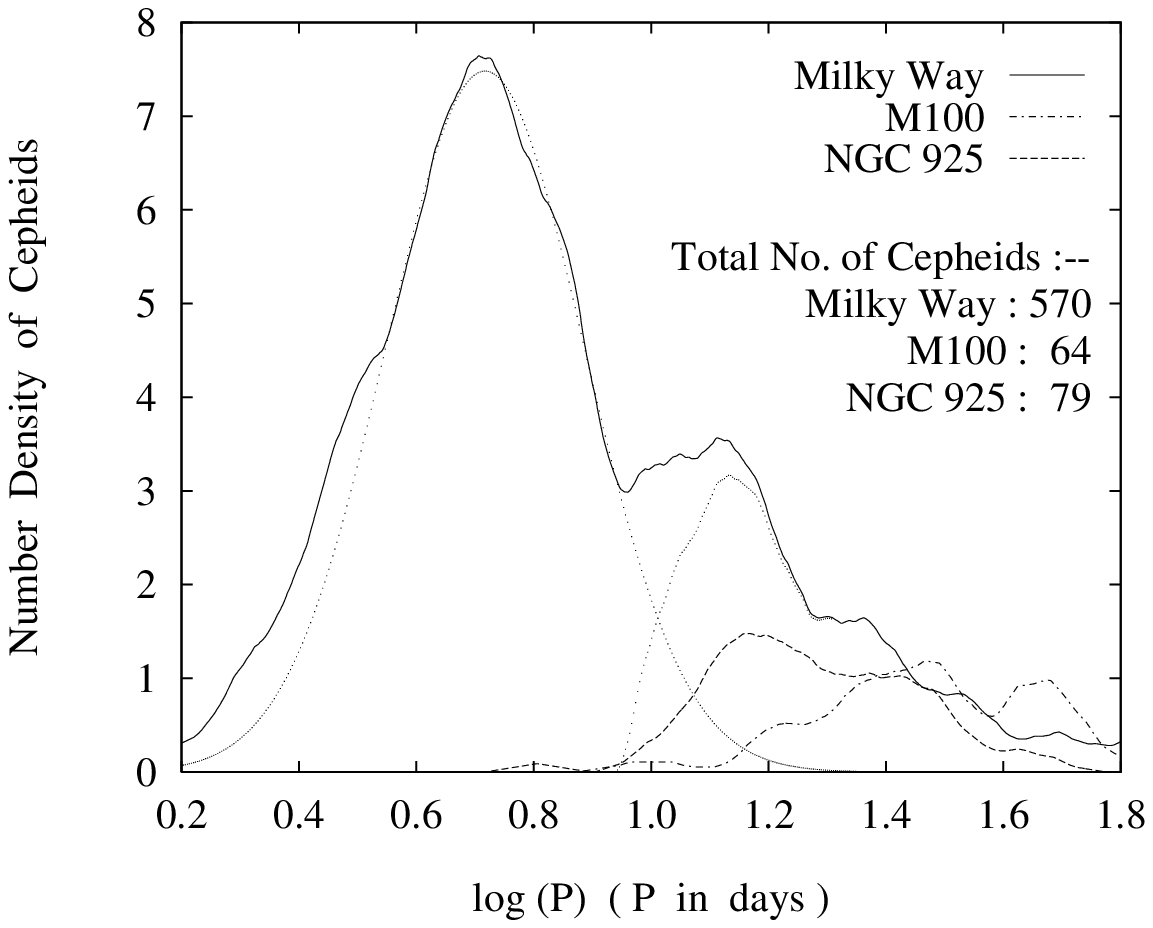}
{The number density distribution against $\lp $ is shown for Cepheids
in the Milky Way, as well as in two other galaxies (observed by HST). A moving
average has been used to generate a smooth curve from discrete observations.
The Galactic sample can be split up into two populations (shown by the
dotted lines), having slight overlap between periods of 9 and 18 days.}
{fig:nod}

We also studied the number density distribution of Cepheids in some of the
galaxies being observed by HST. Two examples (M100 and NGC 925) 
(\cite{ferrarese:96}; \cite{silbermann:96})
are plotted along with the Milky Way distribution in Figure~\ref{fig:nod}. 
We believe that in these galaxies, probably due to flux
limitation, only the higher period Cepheids have been detected. These 
pulsators follow the same number density pattern as shown by Galactic
Cepheid variables. Thus, for the analysis of these extragalactic Cepheids, the
template should be built on Galactic Cepheids having period $\log (P) \geq
1.15$ only, and low period overtone Cepheids should be avoided.
The implications of this preferred selection on the slope of the \plr\ will
be discussed in Paper II.

\section{Synthetic Light Curves}
\label{sec:ligcur}

For extragalactic Cepheids, usually the V (Johnson) band observations 
have good phase sampling, from which generally it is possible to construct
a reliable \lig. However, for observations in other filters, though
equivalent of the I (Cousins) band is a popular choice, often the phase 
coverage is not good, and
it is difficult to obtain a \lig\ with correct values for the various 
parameters like amplitude or peak and mean brightness. 
For Galactic Cepheids, however, in many cases the data is well-sampled
for all the bands, and it is possible to obtain reliable \lig s for the
multi-wavelength bands and the colors as well.
It is instructive as well as useful to compare the general shapes of 
\lig s of different colors like $\BV$, $\vi$ with that of the $V$ \lig\ 
of a particular Galactic Cepheid. We attempt to produce a reasonably accurate
\lig\ for the $\vi$ color from the $V$ \lig\ of a given Cepheid. Although
such attempts to obtain \lig s in other bands from $V$ \lig\ have been
made before (e.g., ~\cite{lst:97}), our approach is based on the phase
difference between the flux variability and the color variability and is
primarily intended to estimate mean $\vi $ and its peak value when the 
I band phase coverage is poor.

We find that the $V$ and the $\vi$ \lig s of the galactic Cepheids 
are very similar, except for a small phase difference between 
the maximum and minimum and the shape of the curves in the descending branch
(i.e., in the part from the minimum magnitude to the maximum). 
Consequently, it is possible to simulate the $\vi$ \lig\ from the $V$ \lig\
in the following way.
We first reduce the $V$ \lig\ to a normalized version, having magnitude of 
$\pm 1$ at minimum and maximum, and phase running from $-0.5$ to $+0.5$, 
with the phase of the maximum brightness (defined as $\vmax $, having the 
minimum magnitude value) being taken as the zero point. The phases 
$\phimax $ and $\phimin $ correspond to the phases where the 
$V$-magnitude value is minimum and maximum (i.e., brightest
and dimmest) respectively. We define
\[ \vmax = V(\phimax) \]
\[ \vmin = V(\phimin) \]
\[ \phitilnew = \phi - \phimax \]
\[ \phiz = \phimin - \phimax \]
\[\phinew = 
           \cases{
			\phitilnew +1, & if $\phitilnew \leq 0.5$ \cr
			\phitilnew -1, & if $\phitilnew >0.5$ \cr
			\phitilnew, & otherwise
                 } \]
\[ \overline{V} = \frac{\vmin + \vmax}{2} \]
\[ \vamp = \vmin - \vmax \]
\beq
V_n(\phinew ) = \frac{V(\phi) - \overline{V}}{\vamp}
\label{eq:v-vn}
\enq

It should be noted that the quantity $\overline{V}$ is only the average of the
$V$ maximum and minimum magnitudes, and should not be confused with the 
flux-averaged mean $V$ magnitude, for which we reserve the symbol $\uncv $.
We note that the following transformation produces a \lig\ which is
very nearly identical to the $\vi$ \lig. 
\beq
\visyn(\phinew ) =
                  \cases{
                          V_{\mathrm{n}}(\phinew \cdot \frac{\phiz}
                          {\phiz-0.04}) & if $-0.5 \leq \phinew <0$\cr
                          V_{\mathrm{n}}(\phinew \cdot \frac{\phiz-0.07}
                          {\phiz}) & if $0 \leq \phinew \leq +0.5$\cr
                        }
\label{eq:vn-visyn}
\enq
We can reconstruct the $\vi$ \lig\ in the correct units of magnitudes from 
this $\visyn$ by making the inverse
transformation of equation~(\ref{eq:v-vn}) as follows:
\beq
\vi (\phinew ) = \overline{\vi } + \viamp \cdot \visyn (\phinew )
\label{eq:visyn-vi}
\enq
where \[ \overline{\vi} = \frac{\vi_{\mathrm{min}} + \vi_{\mathrm{max}}}{2} \]
\[ \viamp = \vi_{\mathrm{min}} - \vi_{\mathrm{max}} \]
Note once again that this average $\overline{\vi }$ is different from the 
flux-averaged $\uncviav$.
The actual \lig s and the synthesized \lig\ for one Cepheid are shown in
Figure~\ref{fig:lig} for comparison. Although the synthetic \lig\ does
not match the actual \lig\ exactly at all phases, it is good enough
to obtain the value of the mean magnitude, $\uncviav $, and 
the amplitude, $\viamp $, within errors of 0.01 mag.

\figone{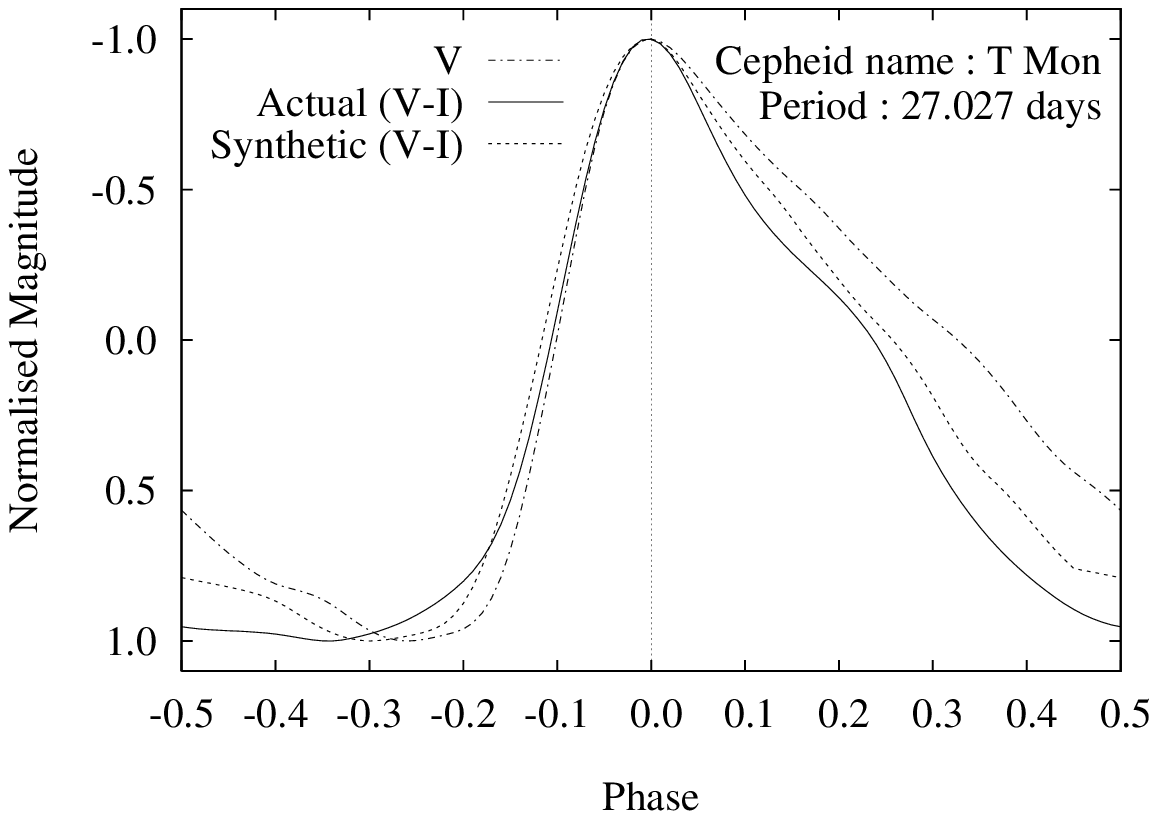}
{Typical \lig s of a Galactic Cepheid are shown. The $\vi $ \lig\
is seen to follow the $V$ \lig\ with a small phase difference. By adjusting
this phase difference, a synthetic \lig\ for $\vi $ may be generated from
the $V $ \lig, which closely simulates the actual $\vi$ \lig.}
{fig:lig}

However, we should stress that all the following analysis for the Galactic
Cepheids was based on the {\em actual} $V$ and $\vi$ \lig s, and not 
on the {\em synthetic} \lig s described above. 
This exercise of synthesizing the $\vi$ \lig\ from the $V$ \lig\ was 
carried out in order to prepare and to scrutinize a template for obtaining 
reasonably accurate \lig s and the associated quantities like amplitude,
mean and peak brightness for data sets where observations in I band
are inadequate to produce independent \lig s for $\vi$. 
This method later proved to be extremely useful for estimating the 
reddening in the HST data for Cepheids, which we shall discuss in Paper II.

\section{Extinction Correction}
\label{sec:extcor}

The Galactic Cepheid variables have generally large extinction (typically
more than 1~mag). The extinction renders the raw colors as well as the 
$V$-magnitude useless even if we know the distance to a Cepheid.
However, considerable work has been carried out to correct for extinction
by studying the spectral characteristics of a background hot star, which
are modified due to absorption in the interstellar medium. In the present
work, we have carried out the extinction correction for the Galactic 
Cepheids by the following formalism. 
According to Cardelli, Clayton \& Mathis (1989),
in grain-dominated interstellar medium, the extinction law can be 
parameterized by the quantity $\rv $ ($= \av/\eBV$).
Based on this $\rv $-dependent extinction law for different
wavelengths, we found that it is possible to derive a
extinction-free quantity in terms of the standard broadband filter colors.
The mean extinction-dependence is of the form
\beq
A_\lambda/\av ~=~ a(\lambda) + b(\lambda)/\rv,
\label{eq:ccm:extlaw}
\enq
where $a(\lambda)$ and $b(\lambda)$ are given polynomials in $\lambda ^{-1}$,
for $1.1~\mu {\mathrm m}^{-1}~\le ~\lambda ^{-1} ~\le 3.3~\mu {\mathrm m}^{-1}$.
From this, the following relations for the reddenings can be derived.
\beq
\left(
\begin{array}{c}
\eUB \\
\eBV \\
\evi \\
\end{array}
\right)
~=~
\left(
\begin{array}{lr}
a({\mathrm U}) - a({\mathrm B}) & \ \ \ b({\mathrm U}) - b({\mathrm B}) \\
a({\mathrm B}) - a({\mathrm V}) & \ \ \ b({\mathrm B}) - b({\mathrm V}) \\
a({\mathrm V}) - a({\mathrm B}) & \ \ \ b({\mathrm V}) - b({\mathrm I}) \\
\end{array}
\right)
\ 
\left(
\begin{array}{c}
\av \\
{\displaystyle \av / \rv} \\
\end{array}
\right)
\label{eq:redd0}
\enq
\noindent
For the standard Johnson-Cousins UBVI filters, the coefficients 
$a(\lambda)$ and $b(\lambda)$ are given by,
\[ \begin{array}{lcrlcr}
a({\mathrm U}) & = & 0.95346 & \ \ \  b({\mathrm U}) & = & 1.90555 \\
a({\mathrm B}) & = & 0.99975 & \ \ \  b({\mathrm B}) & = & 1.00680 \\
a({\mathrm V}) & = & 1.00000 & \ \ \  b({\mathrm V}) & = & 0.00000 \\
a({\mathrm I}) & = & 0.78421 & \ \ \  b({\mathrm I}) & = &-0.56543 \\
\end{array} \]
\noindent
From Equation~\ref{eq:redd0}, we may write 
\beq
\left(
\begin{array}{c}
\eUB \\
\eBV \\
\evi \\
\end{array}
\right)
~=~
\left(
\begin{array}{lr}
-0.04629 & 0.89875 \\
-0.00025 & 1.00680 \\
$\phs$ 0.21579 & 0.56543 \\
\end{array}
\right)
\
\left(
\begin{array}{c}
\av \\
{\displaystyle \av / \rv} \\
\end{array}
\right)
\label{eq:redd1}
\enq

As may be seen from above, the coefficient $a({\mathrm B})-
a({\mathrm V})$ is a very small quantity, and we may approximately 
equate it to zero. 
Hence we define an extinction-independent quantity $Q$ as
\begin{eqnarray}
Q & \simeq &  -[\cUB -\cBV ] - 0.21\, [\cvi - \cBV ] - 0.20\,\cBV \nonumber \\ 
  &    =   & -[\UB -\BV ] - 0.21\, [\vi - \BV ] - 0.20\,\BV 
\label{eq:qdef}
\end{eqnarray}

The stellar atmospheric structure is determined from the value of the surface 
gravity, $g$, the effective temperature, $\teff $ and the
chemical composition (subject to the uncertainties due to mechanical energy
transport and dissipation in the atmosphere). 
The Cepheid instability strip on the HR diagram can be mapped on to a
similar strip in the \gteff\ plane. 
We determine the position of the strip
in the \gteff\ plane from observations and model atmospheres as follows. 
We use synthetic colors computed by Bessell, Castelli, \& Plez (1998) based
on the latest Kurucz model atmospheres to obtain 
theoretical plots of $\cUB$ and $\cvi$ against $\cBV$ for classical Cepheids
within the period range of 15 to 60 days for an assumed \gteff\ relation. 
In this way, a theoretical plot of $Q$ as a function of $\cBV$ is produced.
By demanding that $Q$ be independent of extinction
for each Cepheid, we can find out the amount of reddening $\eBV$ from this plot,
if the Cepheids were to occupy a line in HR diagram. But due to the errors in
observations as well as the finite width of the instability strip, we use
the $Q$ diagram together with the $\cBV$ vs $\lp$ relation to determine the
reddening correction.

\figone{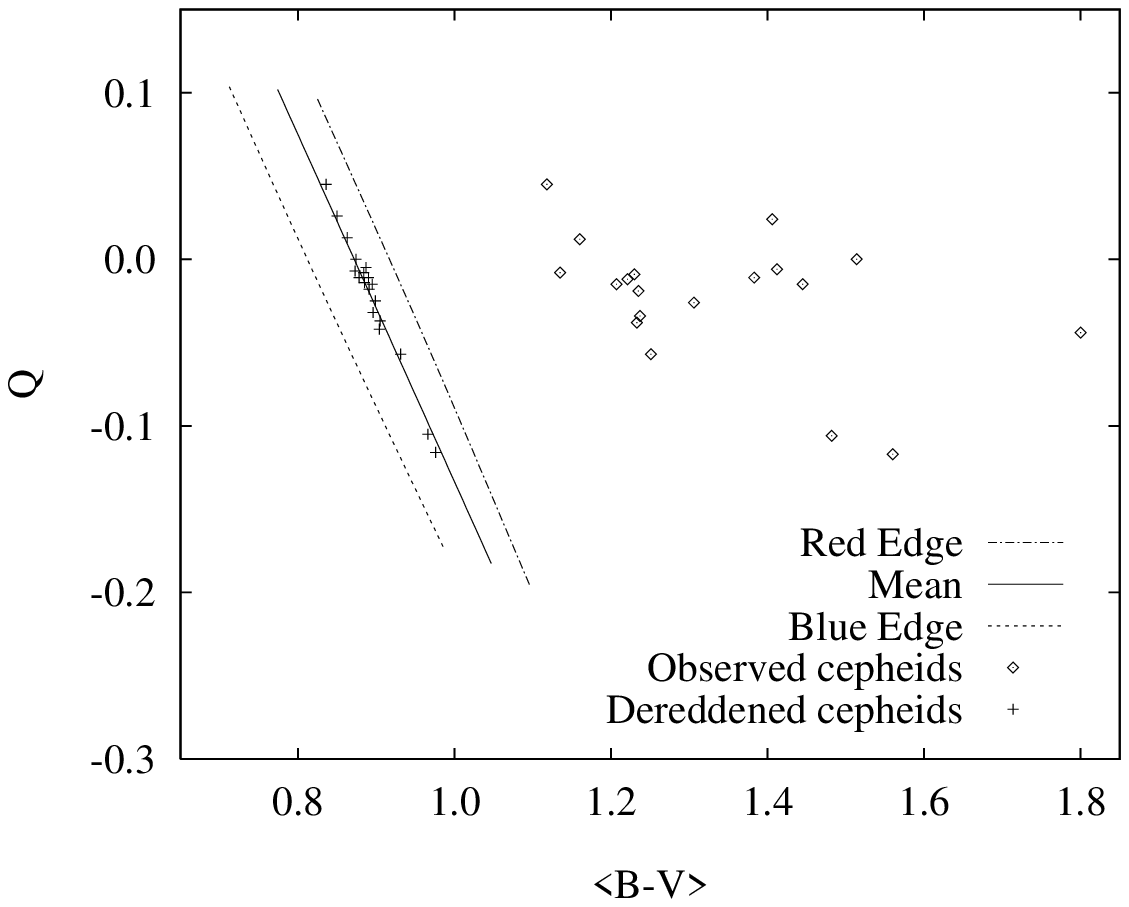}
{The $Q$ Diagram:
The extinction-independent quantity $Q$ is plotted against $\uncBVav$ for the
red edge, the blue edge and the mean of the Cepheid Instability Strip.
The lines are drawn from the model stellar atmospheres on a strip on the
\gteff\ plane. The observed Galactic Cepheids (Table~\ref{tab:cepparam})
whose $\langle B-V \rangle $
are increased due to reddening, should actually occupy positions within
this strip (indicated by ``+'') after dereddening.}
{fig:q_bv}

Since the Cepheid instability occurs within a narrow range of temperature in 
the HR diagram, for a limited range of $\lp $ 
it is reasonable to deduce the existence of a linear 
relationship between $\BVav$ (which is a measure of $\teff$) and $\lp$ (which is
related to the luminosity through the \plr). Indeed, such a relation has been 
found to hold true earlier also (e.g. \cite{fw:87}), albeit with a different
slope compared to what we obtain. We have used a relation of the form
\[ \BVav ~=~a\,\mathrm{log}\, P\,+\,b,         \label{eq:BV-logP} \]
with unknown coefficients $a$ and $b$. On minimizing the total $\chisq$ 
deviation of this relation, along with the $Q$ vs
$\BVav$ graph (Figure~\ref{fig:q_bv}), we obtain the reddening $\eBV $.

After the extinction correction is carried out, we can compare the $\cUB$ and
$\cvi$ for a Cepheid having a specified $\cBV$ with the corresponding
theoretical values obtained from model atmospheres for the assumed strip 
in the \gteff\ plane (Figure~\ref{fig:color:color}). By varying the slope 
and intercept of the strip, we can determine the theoretical
position of the Cepheids in the \gteff\ plane by demanding that the $\cUB$
vs $\cBV$ and $\cvi$ vs $\cBV$ computed from model atmospheres should
individually match the corresponding values derived from the $Q$ diagram.
The strip in the \gteff\ plane which gives the best agreement with the 
observed colors after reddening corrections is shown in Figure~\ref{fig:g_teff}.

\figtwo{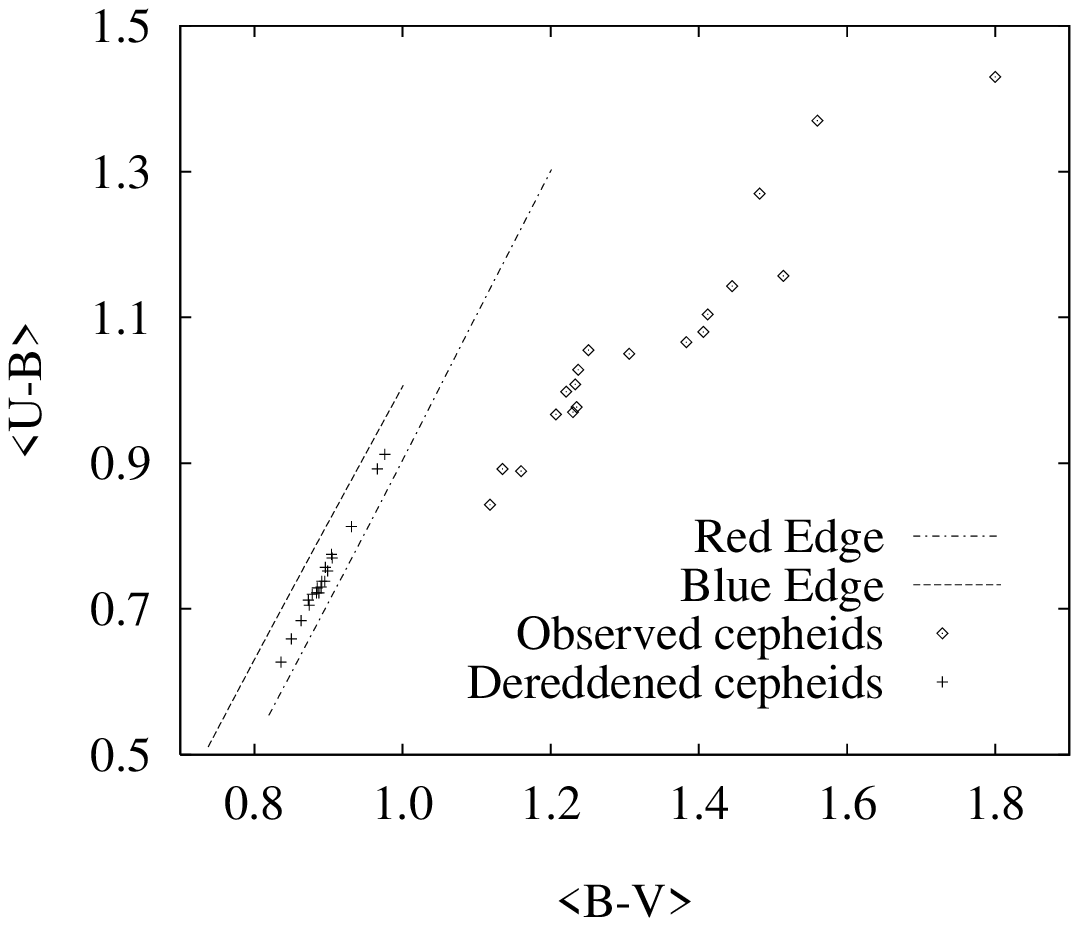}{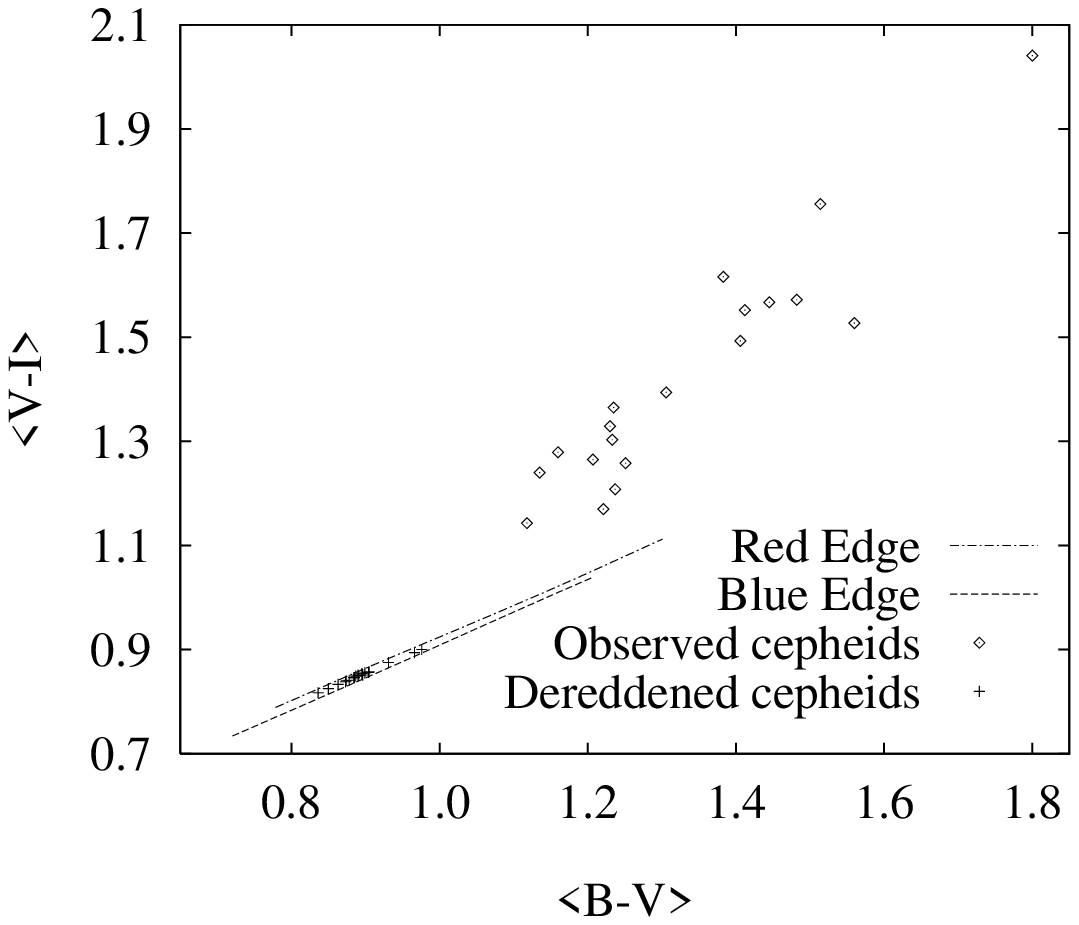}
{The color--color diagrams for the Cepheid instability strip are shown.
The positions of 19 Galactic Cepheids, before and after reddening correction,
are also shown.}
{fig:color:color}
 
\figone{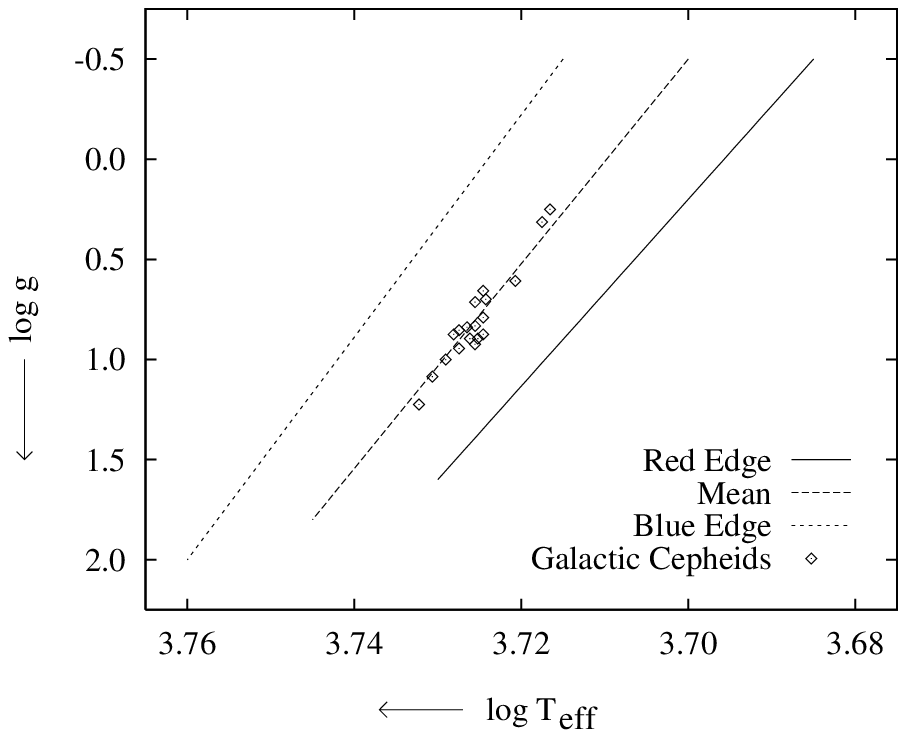}
{The Cepheid instability strip is plotted on the \gteff\ plane.
The positions of 19 Galactic Cepheids, derived from their dereddened colors,
are also shown.}
{fig:g_teff}

Though we started with a fairly large sample, only 25 of the Cepheids
with $\lp \geq 1.15 $ had observations in U, B, V and I bands, showing
tolerable consistency among different observers. Their periods and 
photometric magnitudes (which we obtained by integrating the \lig s) are
listed in Table~\ref{tab:cepparam}. One of the Cepheids, KQ~Sco, had 
very large reddening ($\eBV  > 1$ mag), and since one is not sure about 
the applicability of our statistical formalism to such cases, we chose 
not to include it in our sample.

\begin{table}[h!]
\small
\caption{
Periods and photometric parameters of Galactic Cepheids
\label{tab:cepparam} 
}
\vskip 11pt
\begin{tabular}{lcrccc}
\hline
\hline
\noalign{\vskip 8pt}
{Name} &
{$P$ (days)} &
\multicolumn{1}{c}{$V$} &
{$\UB $} &
{$\BV $} &
{$\vi $}\\
\noalign{\vskip 4pt}
\hline
\noalign{\vskip 8pt}
VW Cen  &  15.036  &  10.231  &   1.080  &   1.406  &   1.493  \nl
SV Mon  &  15.234  &   8.260  &   0.843  &   1.118  &   1.143  \nl
XX Car  &  15.711  &   9.325  &   0.892  &   1.135  &   1.240  \nl
XZ Car  &  16.651  &   8.596  &   1.102  &   1.317  &   1.379  \nl
CD Cyg  &  17.071  &   8.980  &   1.220  &   1.430  &   1.492  \nl
 Y Oph  &  17.127  &   6.124  &   1.066  &   1.383  &   1.616  \nl
SZ Aql  &  17.138  &   8.650  &   1.340  &   1.540  &   1.601  \nl
YZ Car  &  18.165  &   8.714  &   0.889  &   1.160  &   1.279  \nl
VY Car  &  18.912  &   7.482  &   1.028  &   1.237  &   1.208  \nl
RU Sct  &  19.698  &   9.480  &   1.430  &   1.800  &   2.041  \nl
RY Sco  &  20.317  &   8.018  &   1.157  &   1.514  &   1.756  \nl
RZ Vel  &  20.397  &   7.079  &   0.967  &   1.207  &   1.265  \nl
WZ Sgr  &  21.850  &   8.030  &   1.302  &   1.476  &   1.521  \nl
WZ Car  &  23.015  &   9.274  &   0.970  &   1.230  &   1.329  \nl
VZ Pup  &  23.172  &   9.626  &   0.977  &   1.235  &   1.365  \nl
SW Vel  &  23.441  &   8.119  &   1.008  &   1.233  &   1.303  \nl
 X Pup  &  25.965  &   8.513  &   1.050  &   1.306  &   1.394  \nl
 T Mon  &  27.027  &   6.131  &   0.998  &   1.221  &   1.170  \nl
RY Vel  &  28.140  &   8.360  &   1.104  &   1.412  &   1.552  \nl
KQ Sco  &  28.692  &   9.814  &   1.866  &   1.984  &   2.168  \nl
AQ Pup  &  30.072  &   8.681  &   1.143  &   1.445  &   1.567  \nl
KN Cen  &  34.023  &   9.856  &   1.130  &   1.657  &   1.882  \nl
 U Car  &  38.807  &   6.293  &   1.055  &   1.251  &   1.258  \nl
RS Pup  &  41.660  &   6.996  &   1.307  &   1.482  &   1.572  \nl
SV Vul  &  45.103  &   7.240  &   1.370  &   1.560  &   1.527  \nl
\noalign{\vskip 5pt}
\hline
\end{tabular}
\end{table}

We based the selection of the \gteff\ strip on two criteria:
(1) the regained positions of the Cepheids on the 
\gteff\ plane should conform to the original assumed strip;
 (2) theoretical constraint---using the \plr\ (Paper II), we can obtain
the period of a star as a function of $g$ and $\teff $\,:
\beq
0.65\, \lp = -0.5\, \log g - \log \teff + \log Q_{0},
\enq
where $Q_{0} $ is a quantity depending on the effective polytropic index,
which decreases slowly from around 0.025 to 0.015 when the period increases
from 15 to 60 days. Considering the narrow $\teff $
range of the instability strip, the magnitude of the 
slope of $\log g$ against $\lp $ should not exceed 1.35.
For our present analysis, we allow the permissible value of the the slope 
to be in the range 1.2 to 1.35.
We find that at higher $\teff $ values the upper limit of the slope of 
$\log g$ against $\lp $ is exceeded, even for a low inclination of the 
assumed initial strip. 
Further, at higher inclinations, the regained $\teff$ and $g$ values are
completely at variance with the assumed strip (Figure~\ref{fig:g_teff_discr}).
On the other hand, at low $\teff $ for medium to high inclinations
of the strip, although the first condition is satisfied, the slope of
$\log g$ against $\lp $ is far too low. 
If we decrease the inclination
at low $\teff $ the regained positions of the Cepheids do not match the
original strip (see Figure~\ref{fig:g_teff_discr}).
At intermediate temperatures, the limits on the inclination of the strip are
determined by using both the conditions.

\figtwo{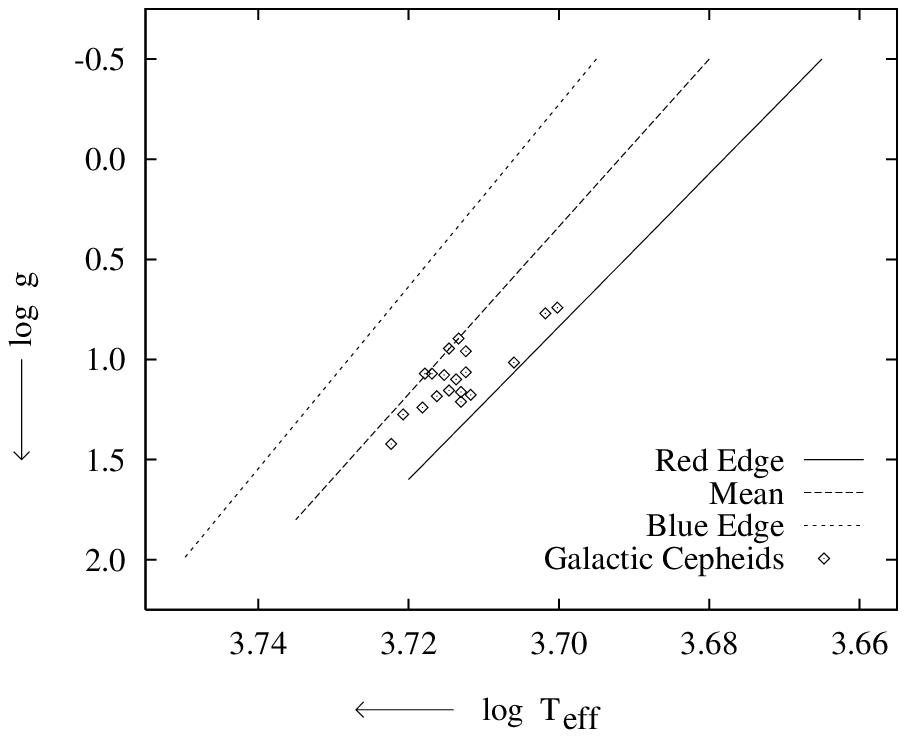}{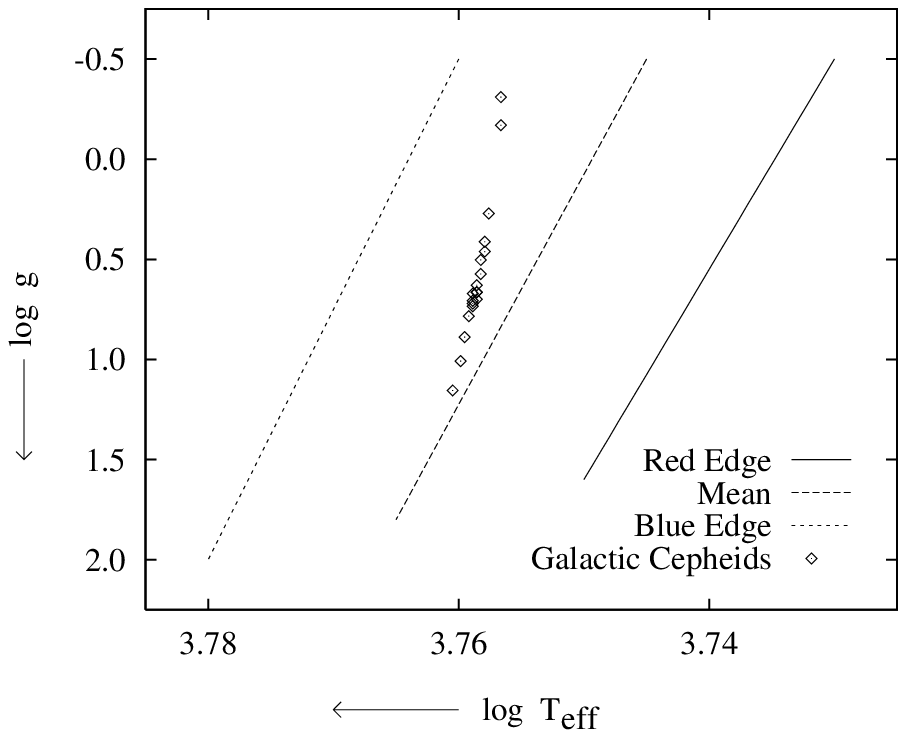}
{The Cepheid instability strip along with the derived positions
of individual stars
is plotted on the \gteff\ plane, for two different choices of $\teff$
vs $\log g $: (a) Low $\teff $ and
low value of $d \log g /d \log \teff $, and (b) High $\teff $ and high
value of $d \log g / d \log \teff $.
These lines provide a limit on the position of the red and blue edge
of the strip.}
{fig:g_teff_discr}

This is the way we converge to the strip shown in Figure~\ref{fig:g_teff}. 
The slope of the $\log g $ vs $\lp $ graph for the accepted strip turns 
out to be $-1.26$ (Figure~\ref{fig:lp_teff_g}), 
which is in agreement with the constraint obtained
on  theoretical grounds. Two typical discrepant strips are shown in 
Figure~\ref{fig:g_teff_discr} for comparison.
It should be noted that the narrow scatter of the points around the mean
line on the \gteff\ plane is an artifact of the statistical methods used
here. Since we have used only two constraints (namely, the $\BVav$ vs $\lp $
relationship and the $Q$ diagram), the minimization procedure causes
the points to be clustered in the final diagrams. The intrinsic width of
the instability strip is not borne out by our converged strip, but 
the mean position of the stars should be considered to be acceptable for 
Cepheids in their core helium burning phase.
It turns out that five among the 24 Cepheids we selected (\namely, XZ~Car,
CD~Cyg, SZ~Aql, WZ~Sgr and KN~Cen) consistently
fall at far too low or high $g$ after the minimization. We interpret it
as an indication that these stars might be in their first or third passage
of the instability strip rather than in the slow blue-ward passage during
core helium burning. Consequently, these five stars were not included in 
our final analysis. Independently such a conclusion was arrived at by
Gieren (1989) for some of these stars, but in two instances (\namely,  VY~Car
and AQ~Pup) our results are different.

\figtwo{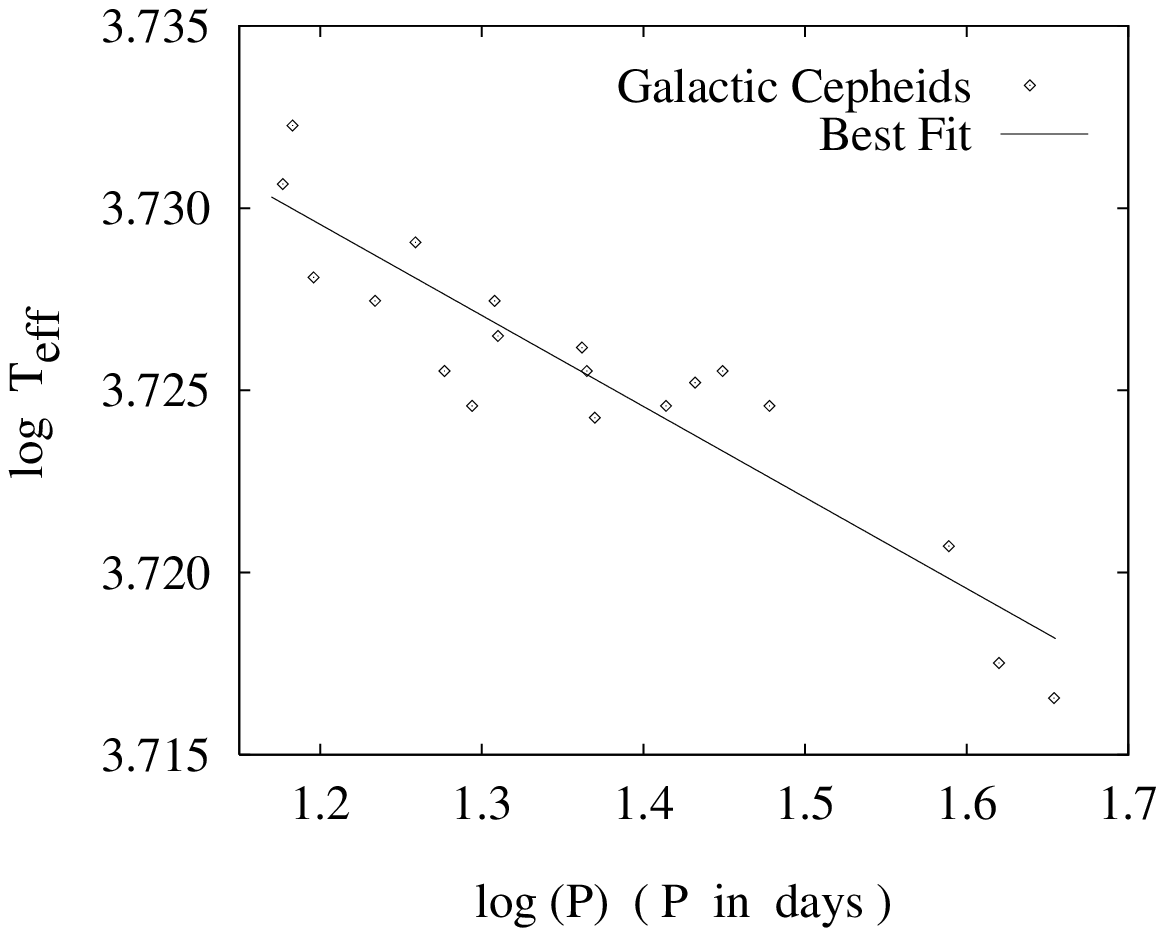}{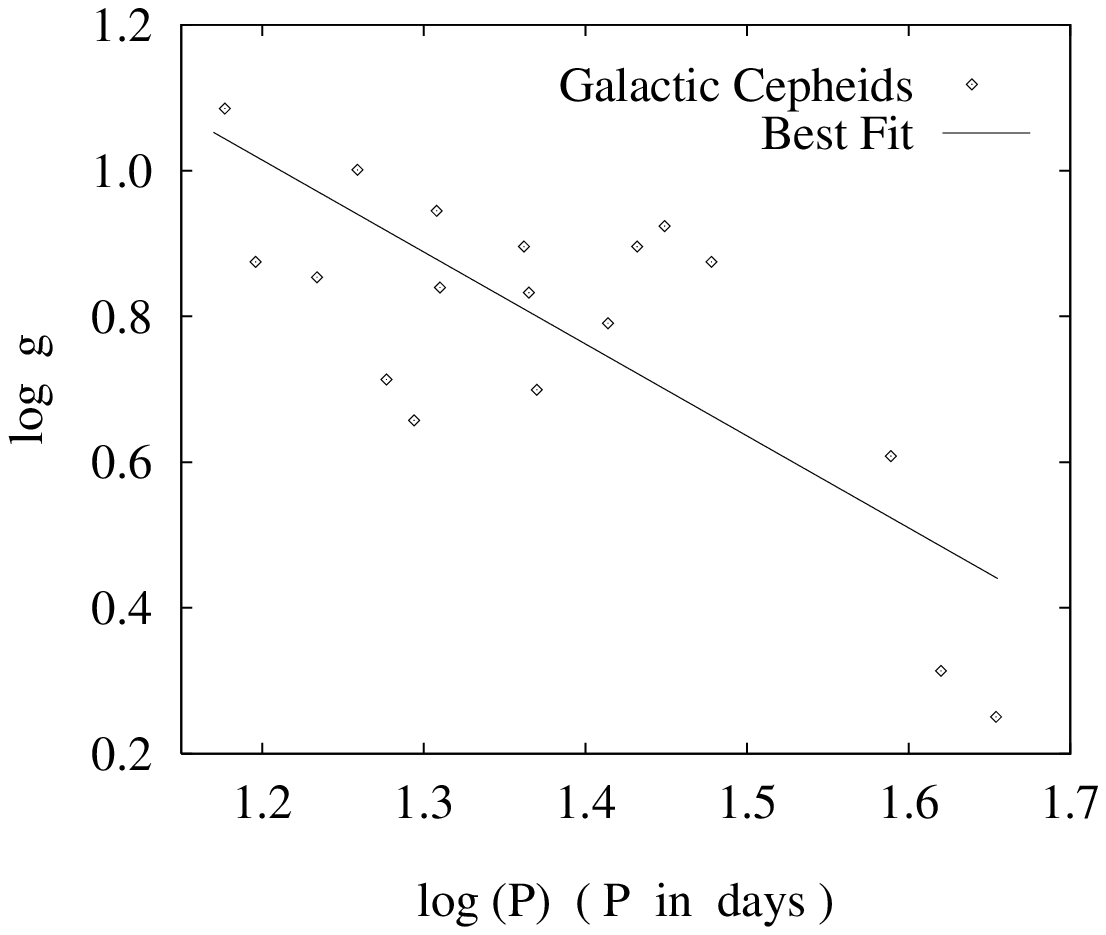}
{Logarithms of the effective temperature, $\teff$, and the
surface gravity, $g$, are plotted against $\lp $ for 19 Galactic Cepheids.
The best fit linear relations are also shown.}
{fig:lp_teff_g}

We obtained a similar strip by combining the evolutionary tracks of 
Bressan \ea (1993) and the luminosity and temperature values given for the 
two edges by Chiosi, Wood and Capitanio (1993).  
On comparison between this strip and that shown in Figure~\ref{fig:g_teff}, 
we find that our values of the surface gravity are lower at the high period 
end than that of Bressan \ea and Chiosi et al. 
This implies that we have a higher luminosity at the
high period end of the strip, which is vindicated by the higher slope of
the \plr, which we shall discuss in Paper II. 
The strip obtained from Bressan \ea and Chiosi \ea runs into inconsistencies 
for the reddening correction technique applied in the present work for Galactic 
Cepheids using model atmospheres (\cite{bcp:98}) when we compare the 
color--color diagrams $\cUB $ vs $\cBV $ and $\cvi $ vs $\cBV $ for the
instability strip with those for the unreddened data.

For the converged instability strip in the \gteff\ plane, the reddening 
$\eBV$ is obtained from the $\chisq$ minimization discussed earlier and
the other reddenings $\eUB$ and $\evi$ are obtained from the color--color 
diagrams.

The reddening-corrected values $\UBav $, $\BVav $, $\viav$ and \mbox{$\vimax$} 
are obtained for each Cepheid. We plot in Figures~\ref{fig:bv_logp} 
and~\ref{fig:virel} the two quantities $\BVav$ and $\viav$ 
against $\lp$ along with the corresponding best fit straight lines given by
the following relations.
\begin{mathletters}
\beq
\BVav = 0.21 \lp + 0.60
\label{eq:bv_logp}
\enq
($\chisq $ per degree of freedom) $= 0.016 $
\beq
\viav = 0.13 \lp + 0.67
\enq
($\chisq $ per degree of freedom) $= 0.008 $\\

We compare our derived reddenings with the values found in the literature,
obtained by other methods (Table~\ref{tab:cepred}). We also compared the
obtained $\BVav $ as a function of $\lp $ with similar relations given
by others. Although the value of the slope of this linear relation is 
generally quoted to be around 0.4 (\cite{fernie:90}; \cite{fc:97}), 
we verified that between the period range 
of $ 1.15 \leq \lp \leq 1.65 $, the slope is indeed 0.22 if we use the data 
from Fernie (1990), which is close to our derived value.
The high value of the slope reported in the literature is valid only if 
low period Cepheids are considered or the unusually large $\BVav $ of
one or two variables which we found to be at a different stage of evolution
are included in the regression. 
This is an example of the differences in various Cepheid 
relations for high and low period ranges (see Section~\ref{sec:modes}).

\begin{table}[!h]
\small
\caption{
Comparison of Reddenings of 19 Galactic Cepheids
\label{tab:cepred}
}
\vskip 11pt
\begin{tabular}{lcccccc}
\hline
\hline
\noalign{\vskip 8pt}
\multicolumn{1}{c}{Name} &
{$\BVav$} &
{}  &
\multicolumn{4}{c}{$\eBV$} \\
\cline{4-7}
{}  &
{This work} &
{}  &
{This work} &
{Fernie\tablenotemark{a}} &
{\ DDO\tablenotemark{b}\ } &
{Caldwell\tablenotemark{c}}\\
\noalign{\vskip 4pt}
\hline
\noalign{\vskip 8pt}
VW Cen & 0.850 & & 0.556 &  0.523 &  0.448 &  0.418 \\
SV Mon & 0.836 & & 0.282 &  0.300 &  0.249 &  0.229 \\
XX Car & 0.873 & & 0.262 &  0.300 &  0.349 &  0.362 \\
 Y Oph & 0.878 & & 0.505 &  0.476 &  0.655 &  0.630 \\
YZ Car & 0.863 & & 0.297 &  0.430 &  0.396 &  0.398 \\
VY Car & 0.896 & & 0.341 &  0.390 &  0.243 &  0.230 \\
RU Sct & 0.904 & & 0.896 &  0.820 &  0.957 &  0.965 \\
RY Sco & 0.874 & & 0.640 &  0.730 &  0.777 &  0.702 \\
RZ Vel & 0.885 & & 0.322 &  0.300 &  0.335 &  0.296 \\
WZ Car & 0.884 & & 0.346 &  0.330 &  0.384 &  0.384 \\
VZ Pup & 0.891 & & 0.344 &  0.350 &  0.471 &  0.478 \\
SW Vel & 0.905 & & 0.328 &  0.340 &  0.349 &  0.356 \\
 X Pup & 0.899 & & 0.407 &  0.410 &  0.443 &  0.417 \\
 T Mon & 0.890 & & 0.331 &  0.370 &  0.209 &  0.172 \\
RY Vel & 0.887 & & 0.525 &  0.240 &  0.562 &  0.558 \\
AQ Pup & 0.895 & & 0.550 &  0.600 &  0.512 &  0.555 \\
 U Car & 0.931 & & 0.320 &  0.350 &  0.283 &  0.277 \\
RS Pup & 0.966 & & 0.516 &  0.640 &  0.446 &  0.484 \\
SV Vul & 0.976 & & 0.584 &  0.580 &  0.570 &  0.431 \\
\noalign{\vskip 5pt}
\hline
\end{tabular}
\tablenotetext{a}{\raggedright From Fernie \& Hube (1968)}
\tablenotetext{b}{From the David Dunlap Observatory
Database of Galactic Classical Cepheids (\cite{fbes:95})}
\tablenotetext{c}{From Caldwell and Coulson (1987)}
\end{table}

\figone{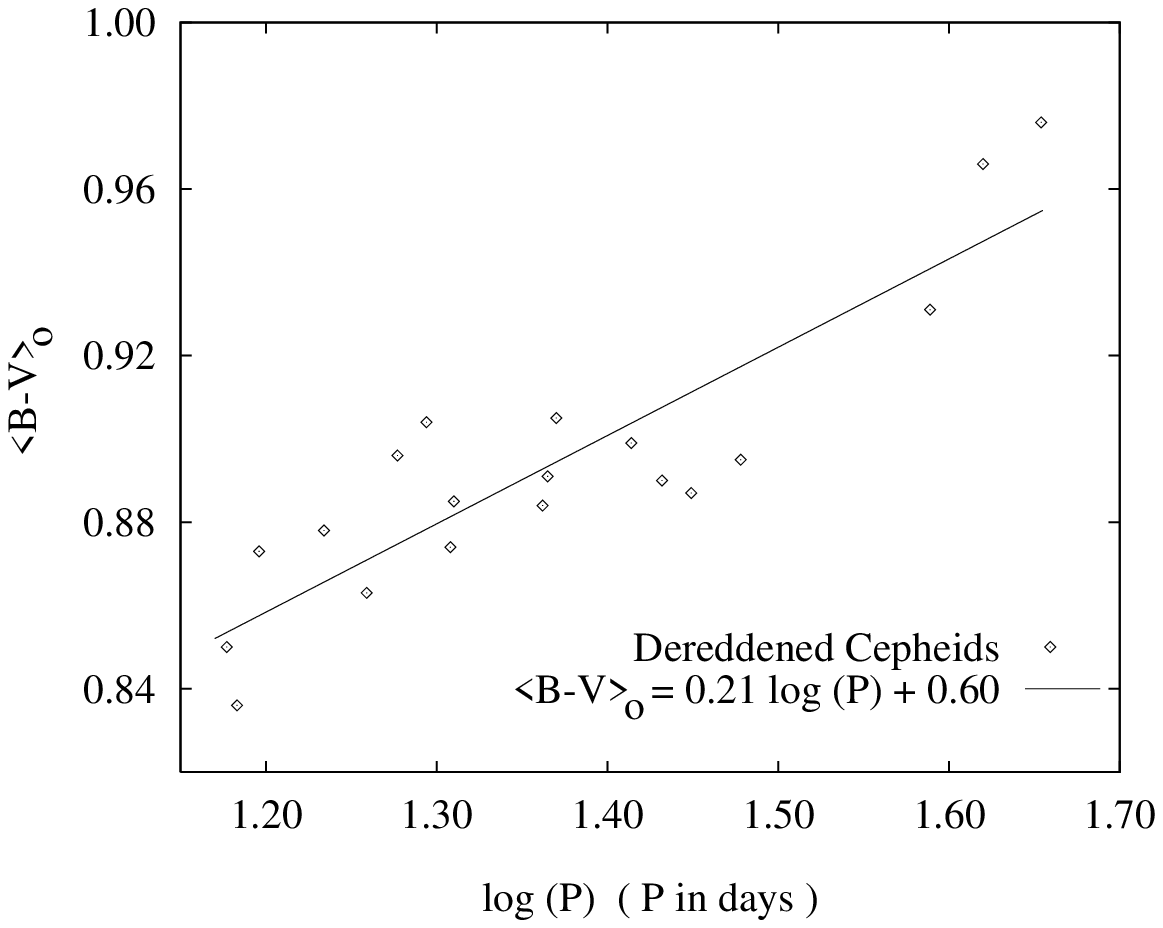}
{The dereddened color $\BVav $ is plotted against $\lp $ for
19 Galactic Cepheids. The best fit line, obtained from $\chisq $ minimization
of this graph along with the $Q$ diagram, is also shown.}
{fig:bv_logp}

We also found that the color $\cvi$ at the brightest phase of the Cepheid, 
i.e., \mbox{$\vimax$}, bears a linear relationship with the amplitude of light
variation, $\vamp$ (Figure~\ref{fig:virel}). (However, due to lack of 
sufficient data, this relationship could not be investigated for 
all the 19 Cepheids.) The derived relation is
\beq
\vimax = -0.28 \vamp + 0.87
\enq
($\chisq $ per degree of freedom) $ = 0.018 $
\end{mathletters}

\figtwo{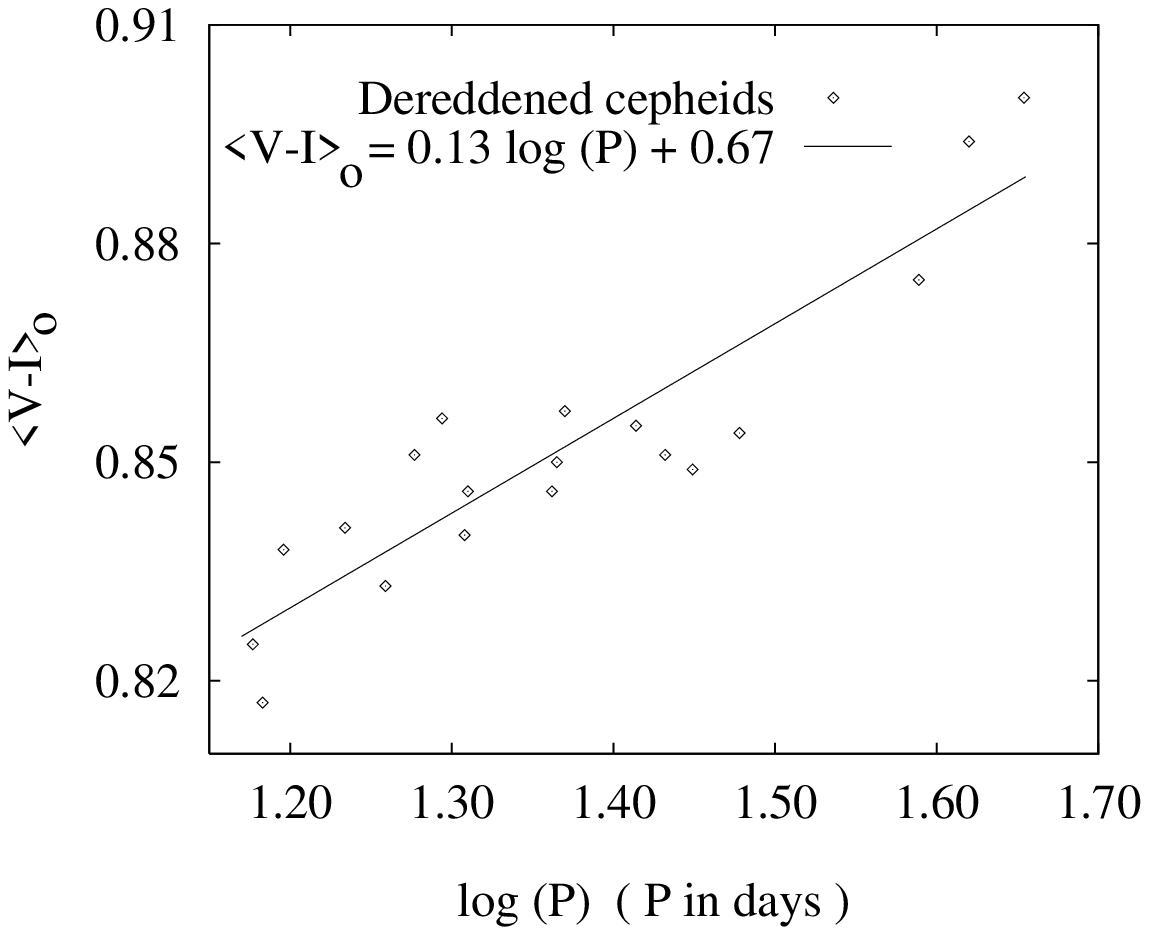}{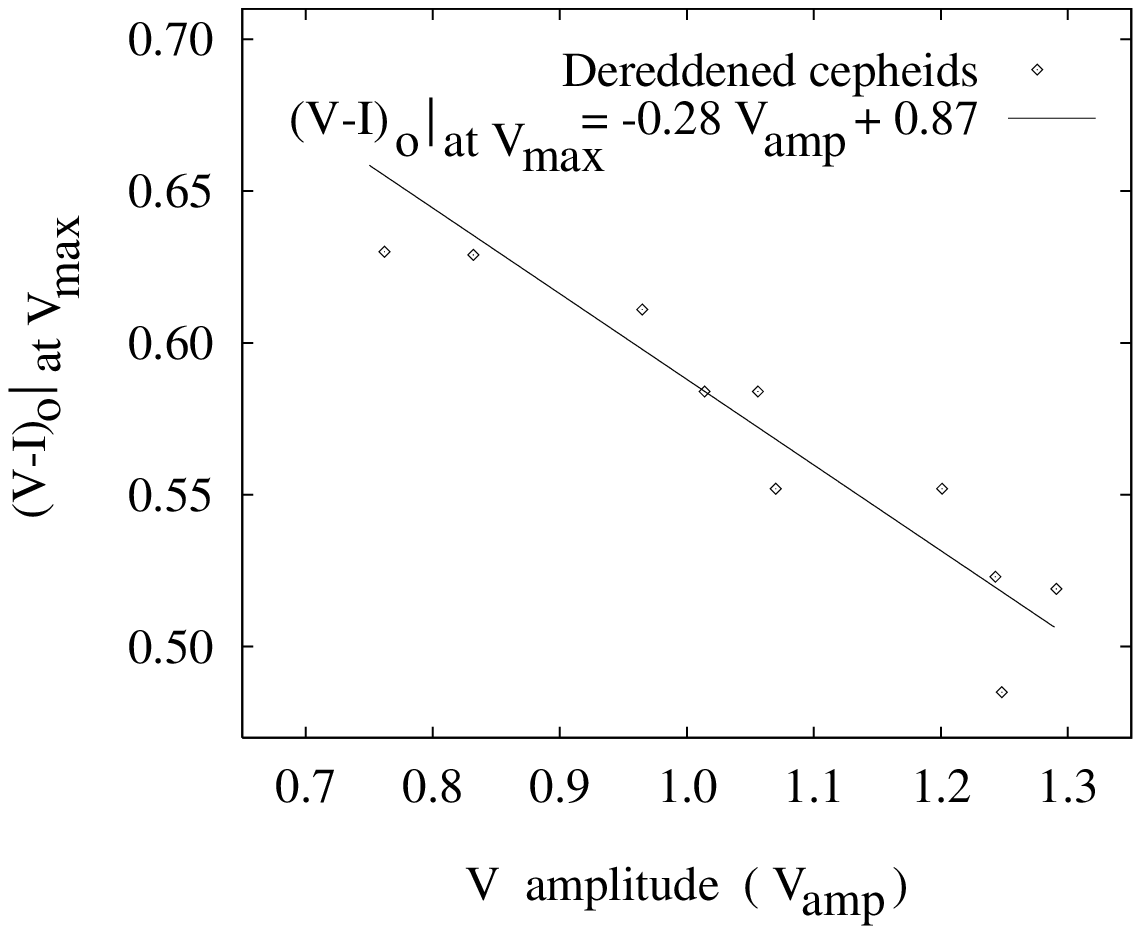}
{The color--period and color--amplitude relations are shown for
Galactic Cepheids.}
{fig:virel}

These latter two relations have been used later for extinction correction
of the HST data on M100 (discussed in Paper II). Since the 
statistics of these relations are constrained by lack of sufficient data, 
these relations should be treated as indicative, and allowance has to
be made for some uncertainty in the values of the coefficients.
Also, since this analysis was restricted to the Galactic classical Cepheid
variables only, the relations we have derived should be considered to be
applicable to populations having chemical composition similar to the average 
over the Galactic distribution. Nevertheless, we are able to establish the 
existence of such linear relationships, and their usefulness for estimating 
dereddened colors will be illustrated in Paper II.

We tried the synthetic colors obtained from model atmospheres by various 
workers, but each appears to have some or other problem while comparing
with observed values of standard stars in the luminosity class and
spectral type of our interest. Finally we selected the latest available
tables from Bessell \ea (1998), but even there they suggested that they
had to rescale their $\UB $ colors to agree with the observed colors of
stars. As discussed later in Section~\ref{sec:limit}, we found that
their unscaled $\UB $ colors fit the Cepheids better, and it is
desirable to increase it by $4 \% $. It is thus evident that the present
model atmospheric colors are not accurate enough for detailed extinction
correction if we have to use the $\UB $ color. Our expressions involving
only $\viav $ and $\vimax $ circumvent this problem to some extent. But
to test the sensitivity of our results to systematic error in the 
synthetic $\UB $ color, we have repeated our analysis with the model
value of $\UB $ increased by $1 \% $ for all $g$ and $\teff $ values.
Though the range of $g$ at which the Cepheid variables of period between
15 and 60 days occupy changes by merely $20 \% $, the converged
expressions for \pca\ relations do not show any
appreciable change. For instance the three relations now become
\begin{mathletters}
\beq
\BVav = 0.21 \lp + 0.62
\label{eq:bv_logp_alt}
\enq
\beq
\viav = 0.13 \lp + 0.69
\enq
\beq
\vimax = -0.28 \vamp + 0.88
\enq
\end{mathletters}
This indicates that the value of $\BV $ or $\vi $ after extinction
correction at typical period of 30 days and 1 mag amplitude remains the
same within $0.1 \% $, provided that we use the same table of model
atmospheres for our calibration of local Cepheids and extinction
correction of extragalactic Cepheids at same range of periods.

\section{Modes of Pulsation in Cepheids}
\label{sec:modes}

Stars of zero-age-main-sequence mass in the range of 7\msun\
to 20~\msun\ traverse the Cepheid instability strip in the 
HR diagram thrice; first, red-wards during the contraction of the helium core, 
next blue-wards during core helium burning, and finally again red-wards 
during the shell-burning phase. The mass, luminosity as well as the 
effective polytropic index of the star will be different for the three 
phases and consequently, the \plr\ is liable to vary slightly depending 
on the evolutionary phase.  However, the duration of time over which a 
star traverses the instability strip is largest for the core helium
burning stage (cf.\ \cite{bit:77}) 
and we should probably expect approximately $90 \%$ of the
Cepheids to be in this stage of evolution. From an inspection of the
evolutionary models it turns out that for stars having constant surface 
temperature, the ratio of the central density to the mean density increases 
rapidly with the mass of the star. Consequently, we should not expect all 
the Cepheids to pulsate in the same mode and obey the same \plr .
It is not surprising that in Section~\ref{sec:number} we found that the number 
distribution as a function of $\lp $ exhibits a two-component structure,
separated at a period of around 10 days.

The differences in the pulsation characteristics between Cepheids
with a range of periods are also borne out from the observed 
properties of Cepheids. The distribution of amplitude of light variation 
during a pulsation cycle against the period of pulsation shows some 
characteristics pointing to the mode of pulsation. 
From the GCVS (\cite{khol:88}), we plot the number density distribution
of Galactic Cepheids against their $V$ amplitudes for the two distinct
period ranges (Figure~\ref{fig:amp_dist}). The number density at each 
value of $V$ amplitude is computed by a moving average method in order to
obtain smooth curves, and the distributions are normalized for direct
comparison. The following features are evident from this comparative study.
For Cepheids of period less than 10 days, the amplitude of oscillation in 
$V$-magnitude varies from less than 0.3 mag to upwards of 1.0 mag with an 
average of approximately 0.7 mag. More than $75 \%$ of these Cepheids have 
amplitude less than 0.85 mag. However, for period greater than 10 days, 
almost all the Galactic Cepheids in the catalogue have amplitudes greater 
than 0.7 mag, with an average of slightly over 1.0 mag.
Only about $25 \% $ Cepheids in this period range have amplitudes less
than 0.85 mag, and $35 \%$ of them have amplitudes higher than 1.1 mag.
The catalogue of Caldwell and Coulson (1987) also indicates similar
features in the distribution of $V$ amplitudes over periods.
Since it is known that the higher tone pulsators have less amplitude 
compared to those pulsating in the fundamental mode,
such a distribution corroborates the hypothesis that lower period Cepheids
are multi-mode pulsators as compared to fundamental mode pulsators which 
generally have higher periods.

\figone{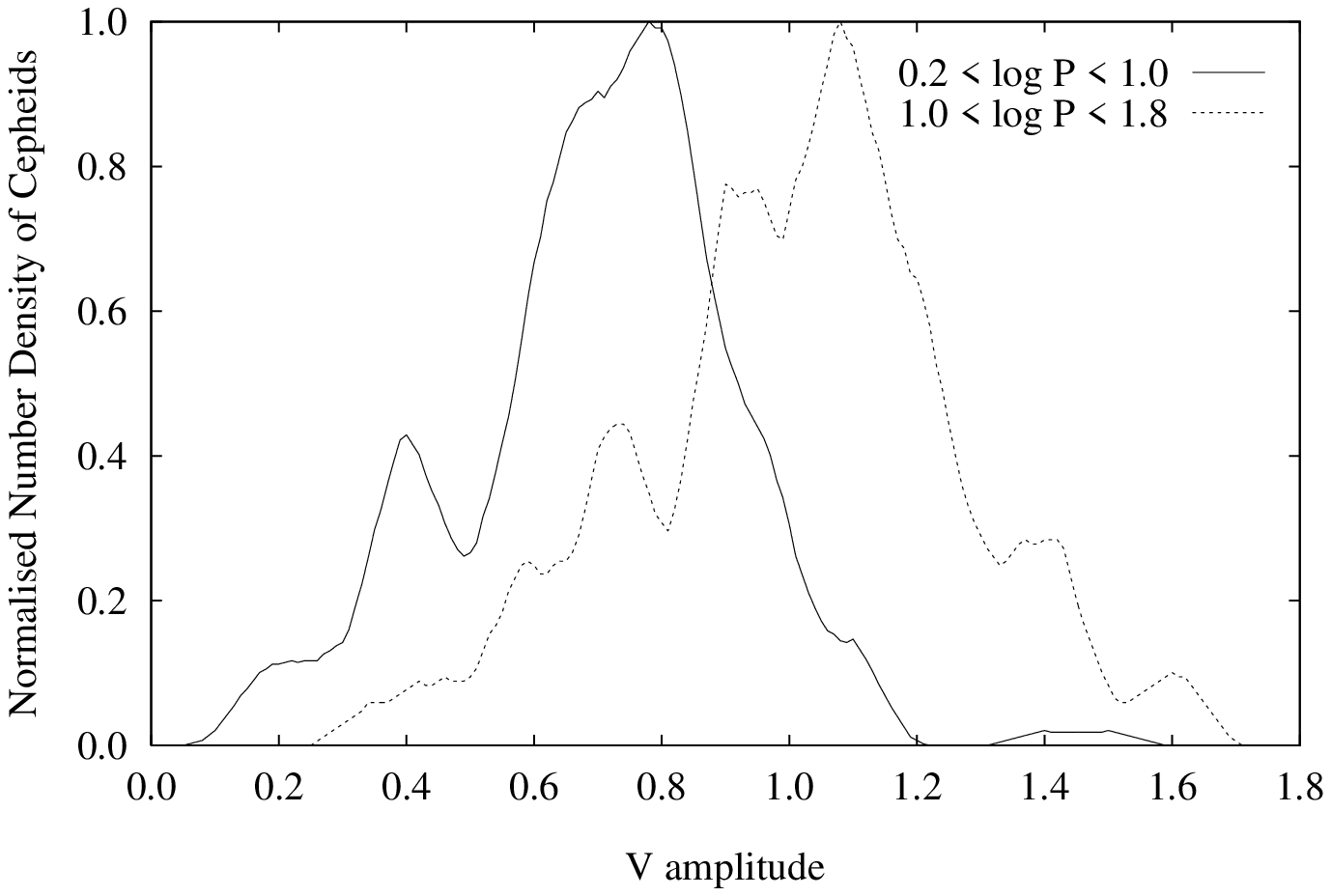}
{The normalized number density of Galactic Cepheids as a function
of their $V$ amplitudes of pulsation are shown for the two ranges of period:
$0.2 < \lp < 1.0$ and $1.0 \leq \lp <1.8$.}
{fig:amp_dist}

Probably the shapes of \lig s of the Cepheids are better indicators of
the modes of pulsation.  At periods between 8--16 days, there are 
multi-mode pulsators called bump Cepheids which are  believed to oscillate 
in fundamental mode and the second harmonic. The bump is considered to be 
due to resonance between these two periods.
At still lower periods, the beat Cepheids are considered to have period of the
fundamental mode  of radial oscillation slightly different from
twice the pulsation period of the second harmonic causing the beat phenomenon.
The MACHO project has revealed many beat Cepheids in LMC 
which pulsate at the fundamental mode and the first overtone, as well as the
first and second overtones (\cite{alcock:95}).

\figone{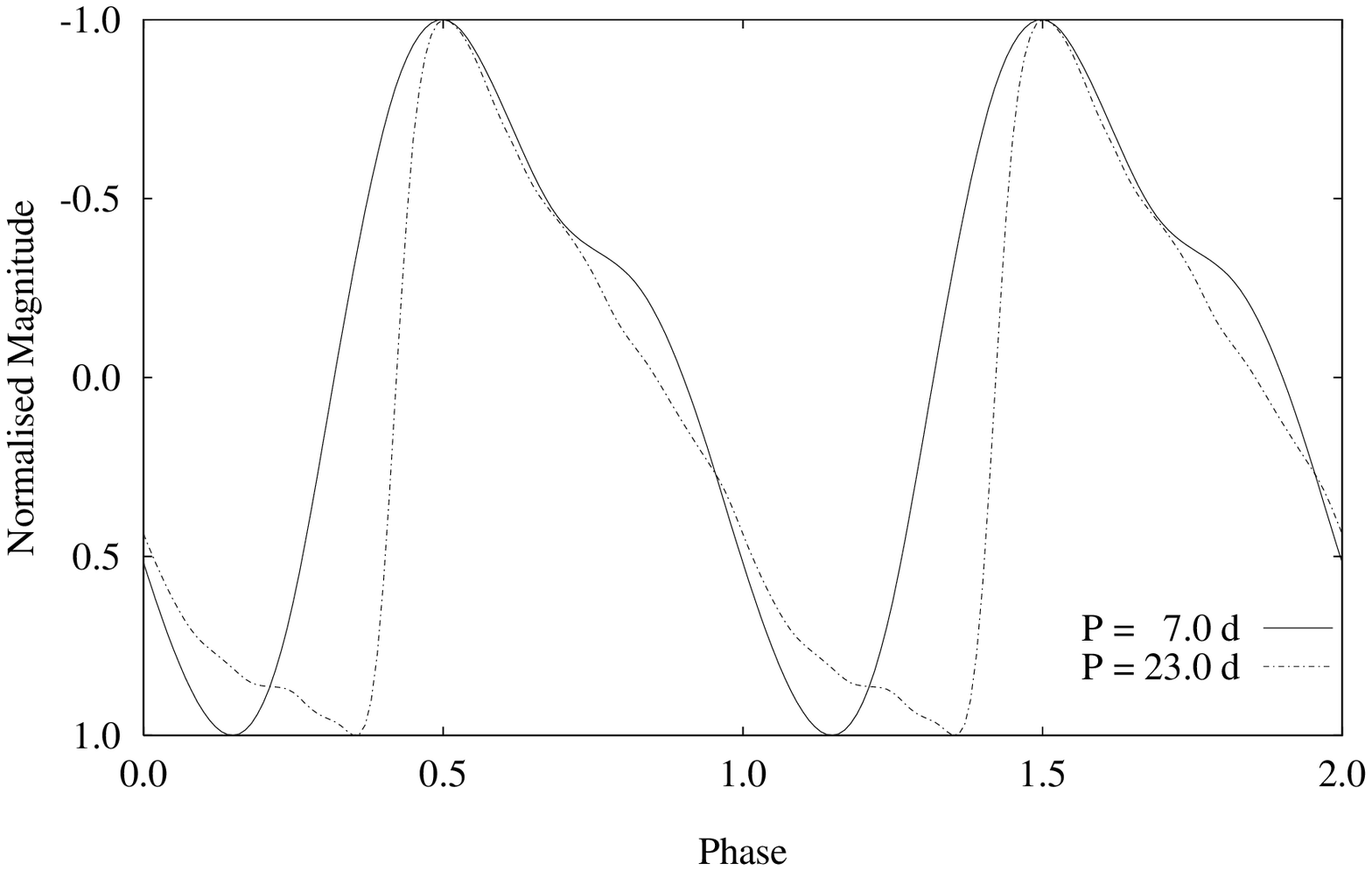}
{Typical \lig s of Galactic Cepheids at periods $7^d$ (V1025~Cyg) and
$23^d$ (WZ~Car) are shown in a normalized form for comparison.}
{fig:lig_comp}

Though there are other low period Cepheids which show smooth \lig s,
there are noticeable differences between the light and color curves of
low and high period pulsators. In Figure~\ref{fig:lig_comp} we have plotted the
$V$ \lig s for a typical 7-day period Cepheid which  shows a smooth
variation of flux, and a 23-day period Cepheid for which the increase of
luminosity with phase is faster and decrease much slower during the
cycle. Also, the smaller period Cepheids often have a ``shoulder'' in the
decreasing branch, which is not detected at higher periods 
(Figure~\ref{fig:lig_comp}). 
The observed minimum in the number of Galactic Cepheids as a function
of their period is believed to be a consequence of mode switching around
this period, thereby producing a dip of approximately 
$\delta \log (P) \sim 0.15 $ which represents the ratio between the
periods of fundamental mode and the first harmonic.
If, indeed, the Cepheids in the entire instability strip were to pulsate
at various modes and their combinations, the Cepheid \plr\ 
should have intrinsic scatter due to contributions from at least 
the following three sources:
\bee
\item 
The ratio of the period of fundamental mode to the second
harmonic of approximately 2 will contribute a scatter of more than 0.3 mag
in the $V$-magnitude.

\item
The dynamical time and effective polytropic index are very different
for Cepheids at the three different phases of evolution, though the 
consequent contribution to the period--$V$-magnitude relation is dependent
on the details of stellar evolution. The scatter in the
period--luminosity diagram due to structural changes during the three
phases cannot therefore be quantified.

\item
For a width of the instability strip of $\delta \log \teff \sim 0.03$,
the spread in the period at constant luminosity will be $\delta \lp \sim 0.09$
and consequent scatter in the $V$-magnitude vs period diagram could be
$\sim 0.3$ mag.

\item
Extraneous factors like metallicity, rotation and diffusion of
elements contribute to the scatter in the \plr\
but no concrete results are available yet.
\ene

The observed scatter in the period--$V$-magnitude relation does not appear
to be substantially larger than 0.3 mag for the Cepheids of high period,
in spite of additional errors due to extinction correction. Consequently,
we are inclined to believe that most of the Cepheids of period larger than 
15 days are likely to be fundamental mode pulsators. Ideally the problem
should be sorted out by an explicit computation of the pulsation
properties of the Cepheids and examination of the stability of the modes.
However, Narasimha (1984) found that the stability of stars in this region
of the HR diagram is crucially dependent on the treatment of the coupling
between pulsation and convection. 
The following discussion is based on the results presented in his thesis.

When the surface temperature of a star is of the order of 8000~K,
all the radial modes are stable, but the first harmonic is only
marginally stable. The stability is a consequence of the weakening of
$\kappa $-mechanism in the helium ionization zone due to a decrease
in the density (discussed in Section~\ref{sec:metal}). 
At lower temperatures when the excitation due to 
$\kappa $-mechanism is stronger, the stability is governed by the onset
of convection in the outer layers where bulk of the dissipation occurs.
Many of the modes can be unstable, but at low enough value of the
acceleration due to gravity at the photosphere, higher harmonics propagate
into the atmosphere, because only the fundamental mode is trapped in the
sub-atmospheric layers.  This follows from the rate of variation of the 
acoustic cutoff frequency in the atmosphere with luminosity along the 
instability strip, where the temperature does not change appreciably
but only mass and luminosity vary. The product ($\Gamma$) of the dynamical
time scale of the star and the acoustic cutoff frequency
varies as $\sqrt{{G \mathcal{M}} \over {R T_{\mathrm{eff}}}}$, where 
$\mathcal{M}$ is the 
stellar mass, $R$ is the radius and $T_{\mathrm{eff}}$ is the effective
temperature. The radius of the star increases by at least a factor
of 5 between periods of 5 and 50 days while the mass of the Cepheid
does not change by more than a factor of 3.
Hence we should expect $\Gamma $ to become less than 1.5 for a Cepheid of
large period during its cool phase of pulsation. Consequently, 
only the fundamental mode can be trapped
in the stellar envelope and the star can pulsate without energy leakage into
the atmosphere. The  higher modes should be traveling up in the
atmosphere and dissipating energy at a rate governed by the local thermal
time scale which is comparable to the pulsation period, and hence they
cannot have well-defined periods. Consequently, irrespective of the
the excitation mechanism for pulsation, the Cepheids of large enough period
should be fundamental mode pulsators. The question is: where does the 
transition from fundamental mode to higher mode or a mixture of modes occur?

The dip in the number versus period of pulsation for the Galactic
Cepheids, at around 10 day period, is indicative of the mode transition 
of the Cepheids. Most of the bump Cepheids occupy this transition region,
between period of 9 to 18 days. 
In order to get a better picture, we appeal to the \gteff\  
plane for the Cepheids, where $g$ is the acceleration due to gravity on the 
surface. Since the stellar atmospheric structure is determined from the 
value of $g$, $\teff $ and the chemical composition (subject to the 
uncertainties due to mechanical energy transport and dissipation in the 
atmosphere), we should, in principle, be able to use the observed colors 
of Cepheids to determine their $g$ and $\teff $.
We have obtained the position of the Cepheids in the \gteff\  plane
by computing their extinction corrected $\cUB $, $\cBV $ and $\cvi $ colors and
have displayed those in Figure~\ref{fig:g_teff} (subject to the qualifications
of Section~\ref{sec:limit}). 
With the help of this figure, we have computed the acoustic cutoff frequency 
for an isothermal atmosphere and the corresponding time scale, 
${\mathcal P}_{\mathrm{ac}}$, given by,
\beq
{\mathcal P}_{\mathrm{ac}} = 2 \pi \sqrt{ H_P \over g} ,
\enq
where $H_P$ is the pressure scale height at the photosphere. In 
Figure~\ref{fig:acoust} we have
shown ${\mathcal P}_{\mathrm{ac}}$ as a function of the period of the Cepheids. 
It is evident that, indeed, ${\mathcal P}_{\mathrm{ac}}$ tends to the pulsation 
period for Cepheids
of period larger than 30 days, while at periods of around 15 days
the fundamental mode as well as first and second harmonic are trapped
below the atmosphere. This is consistent with the observed properties
of bump Cepheids.
Even for the simple isothermal model of the atmosphere of Cepheid
variables of period substantially greater than 20 days, the first harmonic
is likely to propagate into the atmosphere just after the star is hottest
and is about to attain the maximum radius.

\figone{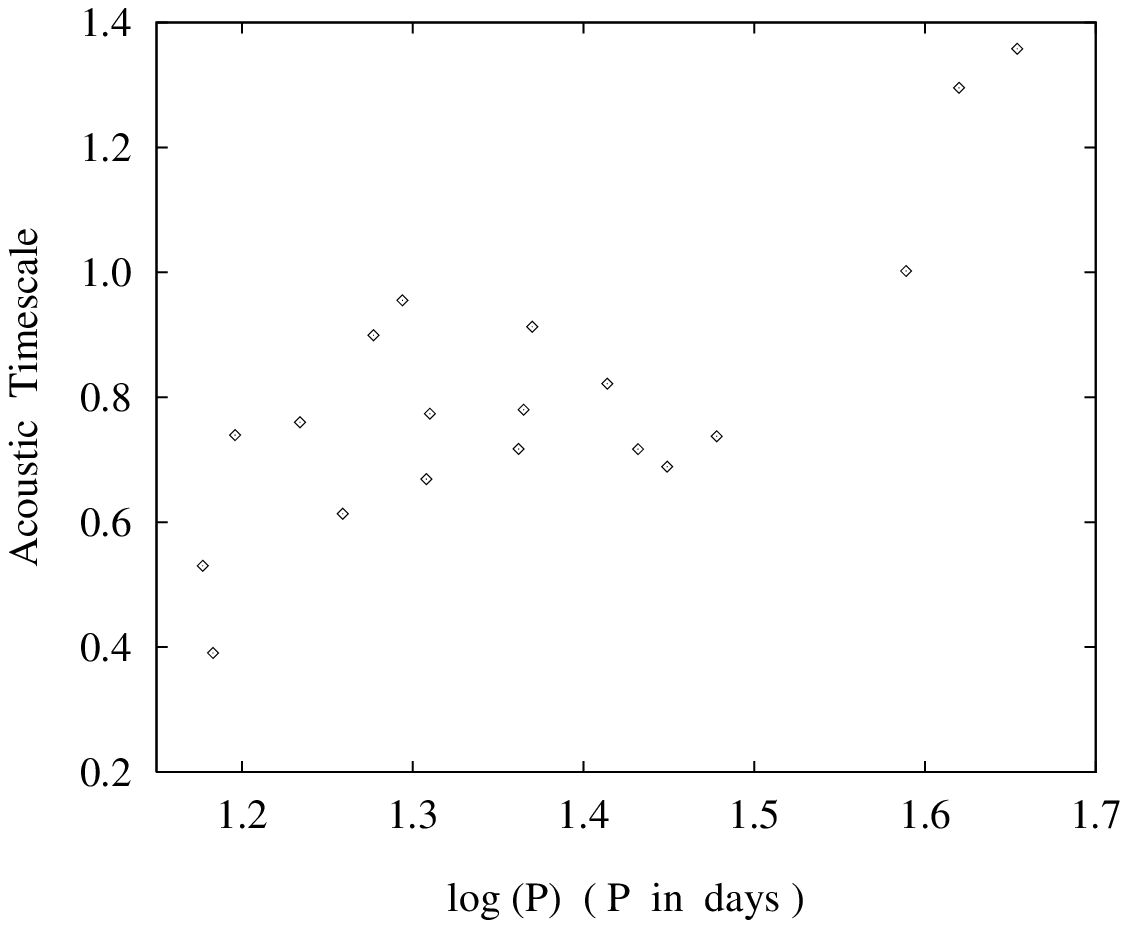}
{The acoustic time scale, ${\mathcal P}_{\mathrm{ac}}$
($ = 2\pi \sqrt{H_P / g}$), is plotted against $\lp $ for 19 Galactic Cepheids.}
{fig:acoust}

A word of caution is required: Cepheids are pulsators of large amplitude
and their atmospheric structure undergoes substantial dynamical change 
during the pulsation
cycle, apart from its being far from isothermal. Specifically, the convective
overshoot into the atmosphere introduces a large input of mechanical energy. 
Our discussion on the propagation of the waves into the atmosphere 
should therefore be treated as qualitative and indicative of the pulsation 
characteristics, rather than as a quantitative result.
The individual points in the graph are not significant because the
assumption that the 19 Cepheids are all fundamental mode pulsators is used
for the computations.

With the help of Figure~\ref{fig:nod} on number density of classical Cepheids 
as a function of their periods, Figure~\ref{fig:amp_dist} on the number 
distribution of Cepheids with respect to their amplitudes of pulsation 
at low and high periods, and 
Figure~\ref{fig:acoust} relating to variation of the acoustic cutoff
frequency with the period of pulsation, we are tempted to conclude that {\em 
Galactic Cepheids of pulsation period greater than 15 days should, in general, 
be considered to be fundamental mode pulsators}. Since this is the preferred
population observed in distant galaxies, their pulsation properties should
be studied independently of those of the low-period oscillators.

As we had discussed in Section~\ref{sec:extcor}, five of the stars were
not considered in the final analysis because their statistically 
inferred positions in the \gteff\ plane were far from the others.
This could be attributed to errors
in the U-band photometry, uncertainty in the computed value of extinction, or,
these Cepheids might not be core helium burning stars. In the latter case,
by determining the position of the star in the color diagram, we indeed 
have a method to identify the core helium burning Cepheids,
which constitute the bulk of the stars in the instability strip.

\section{Cepheid Masses}
\label{sec:mass}

Once we trace the Cepheid instability strip in the \gteff\
plane as a function of the period, we can in principle, use the \plr\
to get an estimate of the mass of the Cepheid
variables independent of the evolutionary or pulsation mass
conventionally used. However, the method has two drawbacks, namely,
(a) the strength of Balmer discontinuity and consequent change in $\UB $
for a  supergiant of spectral type G
depend on both the surface gravity and the hydrogen abundance in
the envelope, and (b) the mapping from $\UB $ to $g$ at fixed $\BV $
is susceptible to large magnifications in the errors in $\UB $ due
to model atmospheric computations.
Nevertheless, since the mass estimates of Cepheid variables could
provide powerful constraints on the evolutionary models of stars of
7\msun\ to 20\msun, 
we have tried to use our results to place the Cepheids 
in the evolutionary tracks adopted from Bressan \ea (1993).

We use the $\teff $ and $g$ values as function of the period
for the 19 Galactic Cepheid variables computed in Section 4
as well as the following \plr\ adopted from Paper II:
\beq
M_V  =  -  3.45 \lp  -  0.79
\enq
The choice of the value of the slope of the \plr\ is based on an analysis of
Cepheids in several other galaxies and is discussed in detail 
in Paper II. The zero point is determined from Hipparcos 
measurement of trigonometric parallaxes of nearby Cepheids (\cite{fc:97}). 
The luminosity at a given period is then computed, using the bolometric 
corrections given by Bessell \ea (1998).

\figtwo{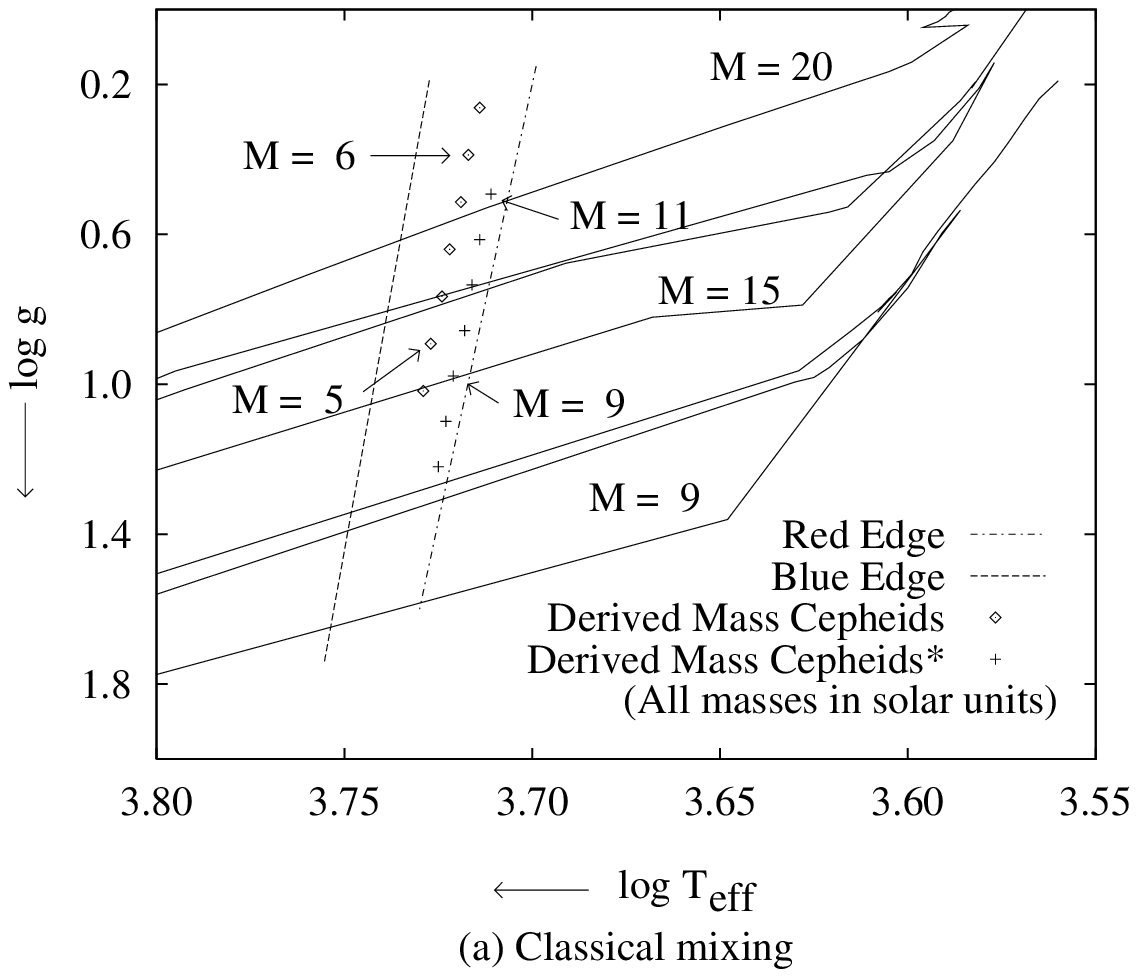}{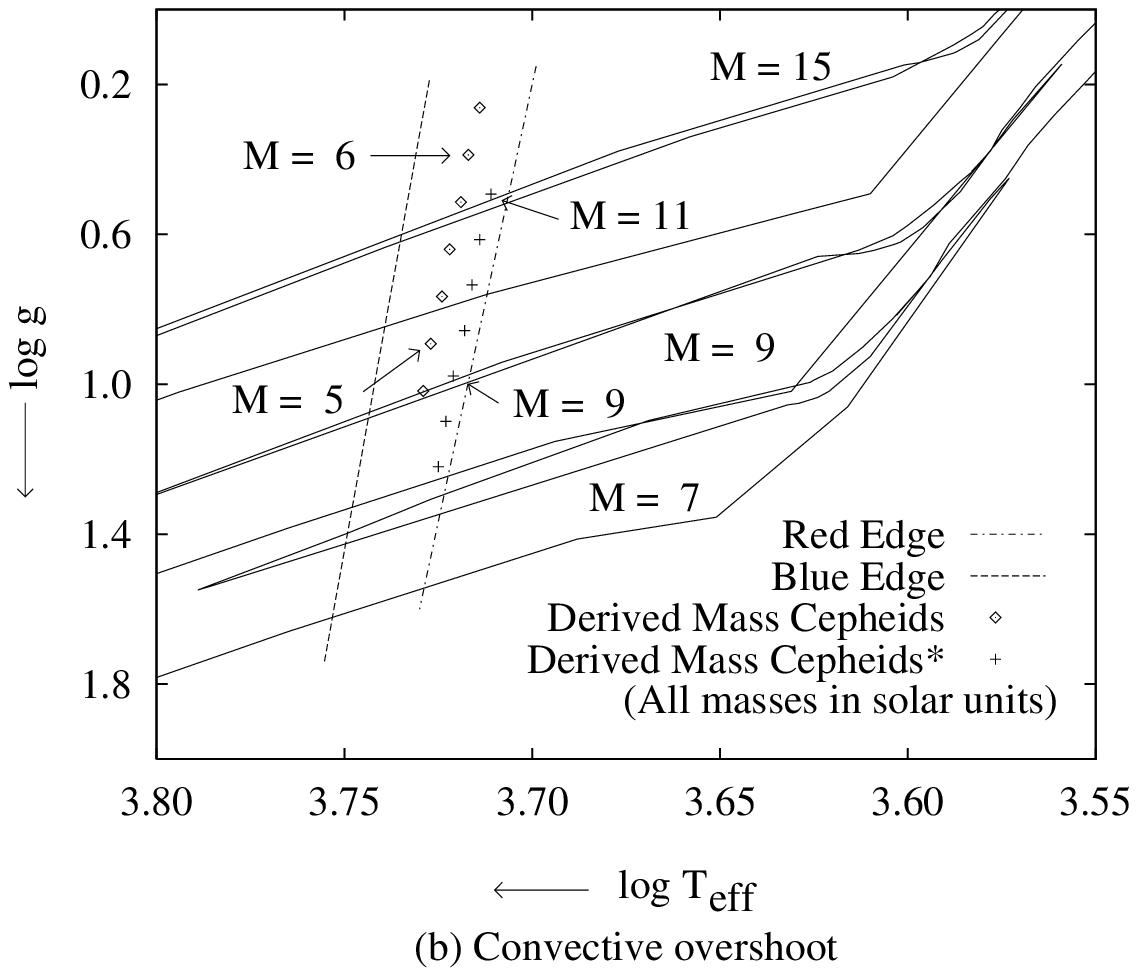}
{The evolutionary tracks of Bressan \ea (1993) have been plotted on
the \gteff\ plane, both for (a) classical mixing and (b) convective
overshoot models. The expected mass of Cepheid variables between
$1.2 \leq \lp \leq 1.8$ based on Figure~\ref{fig:lp_teff_g} are shown.
(* The mass estimates derived with the $\UB $ color for model atmosphere
scaled by 1.02 are also shown with ``+'' symbols.)}
{fig:gt_evol}

We thus estimated the mass of the Cepheid variables along the
instability strip and plotted them in the evolutionary track on a \gteff\
plane in Figure~\ref{fig:gt_evol}.
We compare our derived masses with those obtained from the
evolutionary tracks of Bressan \ea (1993) for models with convective
overshooting as well as without. Evidently, our masses are less by at least
a factor 2 and the extraneous factors responsible to the discrepancy are
discussed above. We can circumvent the problem by assuming that at the
lower end of the period of the Cepheids (15 days), the evolutionary
tracks should be consistent with our computed masses. This is easily
achieved if we adopt the evolutionary tracks with convective overshooting and
use the model atmospheric $\UB $ increased by less than $2\%$. As we had already
demonstrated, this scaling affects the \pca\
relations only in the third decimal. The estimated mass of
around 8.2\msun\ for a period of 15 days is already slightly
higher than the expected ZAMS mass of 8\msun\ for a star
having same $g$ value during its second crossing of the instability strip.
At higher periods, there is a progressively
higher departure between the evolutionary mass and the one we have obtained.
Consequently, even if we assume that for
${\mathcal M} \la$ 9\msun, there is little mass loss up to core
He burning, there should be substantial mass loss and other structural change
modifying $Q_0$ for star of ZAMS mass $\sim$ 15\msun.
Though too much emphasis should not be given to this discrepancy, we
believe that the stars of higher mass ($\ga$ 10\msun)
should be losing
considerable amount of mass during the red giant phase and that their
effective polytropic index as measured from the quantity
$(R/g) (2 \pi/P)^2$ is different from that of low mass Cepheids.

\section{Metallicity Dependency of the Instability Strip}
\label{sec:metal}

Our work does not directly address the problem of metallicity effects on the
Cepheid instability strip. Though ideally pulsation calculations should be 
carried out to address the problem, the coupling between pulsation
and convection in the presence of supersonic convection in an extended
envelope having very low density and relatively small thermal
diffusion time does not provide any easy solution. 
However, in view of the extensive discussions
occurring in the literature, we feel that the following remarks could be of
some relevance:

It has been shown by various workers (cf.~\cite{bit:77}) that for
a fixed temperature, the \plr\ of a Cepheid variable
does not appreciably depend on the chemical composition of the star
(though as we pointed out at the beginning, indirectly there are certain
small secondary effects). Consequently, the two major contributions to
the correction to the \plr\  due to difference in 
chemical composition will be due to
\bee
\item[($i$)] 
shift in the position of the instability strip in the HR diagram,
and
\item[($ii$)]
modifications to the stellar structure during the evolution.
\ene
Specifically, an increase in the temperature of the instability strip at 
constant luminosity  by an amount of $\Delta T$ will produce
a fractional decrease in the period of approximately
$6 \Delta T / T$. (Corrections due to stratification effects, though small,
depend on the model of convection used). But the change in the stellar 
structure as well as evolutionary track crucially depends on the 
mass of the helium core just prior to helium burning. At present the
change in the elemental abundance due to diffusion mechanism as well as 
convective overshooting are poorly understood and consequently, the
detailed evolutionary calculations do not warrant a quantitative 
statement regarding its effect on the Cepheid instability strip.

We can get some insight into the position of the instability strip
without any calculations involving specific models. The overstability of
the radial modes in Cepheid variables is ascribed to the $\kappa$-mechanism. 
For envelopes of supergiants having large acceleration due to gravity
($\log g > 1.2$), the $\kappa$-mechanism operates in the
H and He$^+$ ionization zone at temperature in the range of 6000 to 13000 K
as well as in a narrow He$^{++}$ ionization zone at temperature somewhere
between 29000 K and 50000 K depending on the density of the layer and
extent of the region having density inversion. If we assume that
the H ionization zone determines the blue edge of the instability strip,
beyond which  energization due to $\kappa$-mechanism in the
sub-photospheric layers undergoing hydrogen ionization is more than offset
by the radiative damping in the deeper layers, the
position of this boundary is insensitive to change in hydrogen abundance,
{\em provided that, the metal abundance increases approximately as 
one sixth of the increase in helium abundance in the envelope}. 
For helium abundance of 0.3 by mass fraction the logarithm of temperature
of the blue edge, $\log T_{\mathrm{blue}}$ decreases by an amount of 0.005 
for an increase in the metal abundance of 0.008
for fixed surface gravity.
The blue edge determined in this form, using the opacity tables provided
by Cox and Tabor (1976) as well as Cox (1991) 
is approximately along the line $\log \teff = 3.759$, $\log g = 1.0$
and $\log \teff = 3.787$, $\log g = 2.0$ for metal abundance
$Z = 0.024$ and helium abundance $Y = 0.30$ by mass fraction when envelope
models were run according to the prescriptions described in Narasimha (1984).

At low values of the surface gravity, hydrogen ionization zone alone does
not determine the blue edge of the strip. But since the convective velocity
is model-dependent and is a few tens of km\,s$^{-1}$, the structure of the
envelope itself cannot be specified with any reliability unlike in the case of 
low luminosity stars. Still, if we use the strength of $\kappa$-mechanism 
both in the hydrogen and helium ionization zones
as a measure of the pulsation instability, the position of the blue
edge can be extended to surface gravity of the order of 
$\log g = 0.4$ from $\log g > 1$.

The blue edge determined in this form is in agreement with the extreme
blue boundary obtained from the analysis of Galactic Cepheids 
(Section~\ref{sec:extcor}).
Consequently, we are tempted to say that {\em the Cepheid instability
strip is approximately 0.08 wide in $\log \teff$ for fixed surface
gravity and the position is weakly dependent on chemical composition if
both helium and metal abundance in the zero age main sequence model
are increased simultaneously}. 
However, it would be desirable to analyze colors of
Cepheids of known chemical composition in the envelope using reliable
atmospheric models to determine the sensitivity of the mean position of
the instability strip as a function of chemical composition.

\section{Main Limitations of the Work}
\label{sec:limit}

In this work we have presented an analysis of observations relating
to Cepheid variables compiled from various sources, coupled with 
theoretical considerations and simple calculations based on envelope models. 
Note the following limitations while using these results for calibrating 
the Cepheid instability strip:

\noindent
1. The data we have used is culled from a number of sources. Note that
the data have been compiled over
more than thirty years, during which time the efficiency of detectors in U 
band has evolved and various workers may not have always taken into
account the changes in the definition of different I band filters consistently.
Though we have tried to cross-check the data and taken a sample which we 
consider to be trustworthy, it would be desirable to take a homogeneous sample
from a single source for the kind of work we have carried out.
Clearly, a systematic zero point error of 0.05 magnitude between data from
two groups would lead to misleading final results.
The size of our final sample---namely, 24 Cepheids having period 
greater than 15 days, 5 stars of which are
at a different evolutionary phase---turns out to be small and could cause
considerable random error in the relations we derived.

\noindent
2. We adopted the model atmospheric data on the color vs $\teff $ and $g$
from Bessell \ea (1998) after trying a few other available tables.
However, they had to rescale their $\UB $ to match the observational data.
We found that for the stars of our interest, namely, late F to early K
supergiants, their original $\UB $ is better than the rescaled data and that
by increasing the value by approximately $4\%$, we can get better agreement
for the Cepheids. Fortunately, a change of $1\%$ in their $\UB $ 
shifts the Cepheid instability strip by only 0.003 in $\log \teff $ for a
fixed $g$ value and the slopes and intercepts of the \pca\ 
relations change only in the third decimal, which is less than
the typical errors in the slope. Nevertheless, it is clear that
{\em if we calibrate a set of data on Cepheids using one model atmospheric
scheme, we have to use the same scheme for another set of observations
used to find the distance to a source or other characteristics of Cepheids}.

But we would like to stress that
\bee
\item[a.]
Our  $\log \teff $ at fixed period for the mean position of the 
instability strip could have an error of possibly more than 0.01 due to 
the above two limitations.
\item[b.]
The value of the mean $g$ for a fixed period of the Cepheid
is more sensitive to the errors caused by $\UB $. A change in the zero point of
$\UB $ by $1\% $ changes the derived mass by $30\% $. The fact that we have
obtained a reasonable (though too low) mass for the Cepheids as well as
extinction consistent with what is available in the literature indicates
that probably our choice of Cepheid data as well as the model atmospheric
tables might not be bad.
\ene

\noindent
3. We have not carried out any pulsation computations with full stellar
models obtained from evolutionary sequence nor studied the effects of
convection. Consequently, our statements on modes of pulsation of
Cepheid variables having high period or the theoretical position
of the instability strip should be treated as empirical.

\noindent
4. We have been unable to get good enough sample of data for Cepheids in
LMC or other galaxy where extinction is low and metallicity is very
different from that of the Galaxy. To get an idea of the systematic
errors in the calibration of the Cepheid \plr,
it is very desirable to carry out the required observations and analysis.

\section{Conclusions}
\label{sec:concl}

The principal outcome of the present work may be summarized as follows:
\bei

\item[*]
It appears that generally Galactic Cepheid variables of periods greater
than 15 days are fundamental mode radial pulsators, 
while those with lower periods could be mixed mode pulsators or might even be 
oscillating at higher overtones. 

\item[*]
A prescription to estimate the Cepheid $\vi$ \lig\ from the observed $V$ 
\lig\ could be obtained based on the phase shift of \mbox{$\uncvimin$} with
respect to the phase of $\vmin $, whenever the data for the former
is insufficient to draw an independent \lig. 

\item[*]
Relations between pulsation period
and color ($\BV$,$\vi$), as well as amplitude of light variation in V band
and color at brightest phase, \mbox{$\vimax$} can be derived from the observed 
\lig s. The extinction correction can be carried
out either by using mean $\UB $, $\BV $ and $\vi $ as functions of
period or from the mean $\vi $ and its value at the brightest phase of
pulsation. 

\item[*]
An estimation of Cepheid masses as function of period,
is given by determining the instability strip in the \gteff\ plane and using
model atmospheres to argue that for stars with ZAMS mass greater than 10\msun, 
mass loss and consequent structural changes cannot be neglected.

\item[*]
The metallicity effects on the Cepheid instability strip are estimated
and the position of the blue edge of the strip is obtained from the strength 
of the $\kappa$-mechanism to find that the metallicity effect 
on the \plr\ is not important if helium abundance increases with
increasing metal abundance.

\eni
\noindent
The limitations of the work should be noted while using our expressions and
techniques for calibration of the Cepheid instability strip.

\acknowledgements
We are grateful to Fiorella Castelli for providing us the latest model 
atmospheres (ATLAS9) in an electronic format, and to
S. M. Chitre for his critical comments on the manuscript.
We acknowledge support from
the Indo-French Center for the Promotion of Advanced Research
(Project 1410-2).

\end{document}